\documentclass[twocolumn]{aastex63}

\usepackage{epsfig,natbib}
\usepackage{graphicx}
\usepackage{amsmath}
\usepackage{booktabs}
\usepackage{longtable}
\usepackage{xspace}
\usepackage[T1]{fontenc}
\usepackage{lipsum}
\usepackage{verbatim}

\shortauthors{MacDougall et al.}
\shorttitle{TKS System Properties}
\begin{document}

\title{The TESS-Keck Survey. XV. Precise Properties of 108 TESS Planets and Their Host Stars}

\author[0000-0003-2562-9043]{Mason G.\ MacDougall}
\affiliation{Department of Physics \& Astronomy, University of California Los Angeles, Los Angeles, CA 90095, USA}

\author[0000-0003-0967-2893]{Erik A.\ Petigura}
\affiliation{Department of Physics \& Astronomy, University of California Los Angeles, Los Angeles, CA 90095, USA}

\author[0000-0003-0742-1660]{Gregory J. Gilbert}
\affiliation{Department of Physics \& Astronomy, University of California Los Angeles, Los Angeles, CA 90095, USA}


\author[0000-0002-9751-2664]{Isabel Angelo}
\affiliation{Department of Physics \& Astronomy, University of California Los Angeles, Los Angeles, CA 90095, USA}

\author[0000-0002-7030-9519]{Natalie M. Batalha}
\affiliation{Department of Astronomy and Astrophysics, University of California Santa Cruz, Santa Cruz, CA 95060, USA}

\author[0000-0001-7708-2364]{Corey Beard}
\affiliation{Department of Physics \& Astronomy, University of California Irvine, Irvine, CA 92697, USA}

\author[0000-0003-0012-9093]{Aida Behmard}
\altaffiliation{NSF Graduate Research Fellow}
\affiliation{Division of Geological and Planetary Science, California Institute of Technology, Pasadena, CA 91125, USA}

\author[0000-0002-3199-2888]{Sarah Blunt}
\altaffiliation{NSF Graduate Research Fellow}
\affiliation{Department of Astronomy, California Institute of Technology, Pasadena, CA 91125, USA}

\author[0000-0002-4480-310X]{Casey Brinkman}
\altaffiliation{NSF Graduate Research Fellow}
\affiliation{Institute for Astronomy, University of Hawai`i, Honolulu, HI 96822, USA}

\author[0000-0003-1125-2564]{Ashley Chontos}
\altaffiliation{Henry Norris Russell Fellow}
\affiliation{Department of Astrophysical Sciences, 4 Ivy Lane, Princeton University, Princeton, NJ 08544 USA}
\affiliation{Institute for Astronomy, University of Hawai`i, Honolulu, HI 96822, USA}

\author[0000-0002-1835-1891]{Ian J. M. Crossfield}
\affiliation{Department of Physics \& Astronomy, University of Kansas, Lawrence, KS 66045, USA}

\author[0000-0002-8958-0683]{Fei Dai}
\affiliation{Division of Geological and Planetary Science, California Institute of Technology, Pasadena, CA 91125, USA}

\author[0000-0002-4297-5506]{Paul A.\ Dalba}
\altaffiliation{Heising-Simons 51 Pegasi b Postdoctoral Fellow}
\affiliation{Department of Astronomy and Astrophysics, University of California Santa Cruz, Santa Cruz, CA 95060, USA}
\affiliation{SETI Institute, Carl Sagan Center, 339 Bernardo Ave, Suite 200, Mountain View, CA 94043, USA}

\author[0000-0001-8189-0233]{Courtney Dressing}
\affiliation{Department of Astronomy, University of California Berkeley, Berkeley, CA 94720, USA}

\author[0000-0002-3551-279X]{Tara Fetherolf}
\altaffiliation{UC Chancellor's Fellow}
\affiliation{Department of Earth and Planetary Sciences, University of California, Riverside, CA 92521, USA}

\author[0000-0003-3504-5316]{Benjamin Fulton}
\affiliation{NASA Exoplanet Science Institute/Caltech-IPAC, Pasadena, CA 91125, USA}

\author[0000-0002-8965-3969]{Steven Giacalone}
\affiliation{Department of Astronomy, University of California Berkeley, Berkeley, CA 94720, USA}

\author[0000-0002-0139-4756]{Michelle L. Hill}
\affiliation{Department of Earth and Planetary Sciences, University of California, Riverside, CA 92521, USA}

\author[0000-0002-5034-9476]{Rae Holcomb}
\affiliation{Department of Physics \& Astronomy, University of California Irvine, Irvine, CA 92697, USA}

\author[0000-0001-8638-0320]{Andrew W.\ Howard}
\affiliation{Department of Astronomy, California Institute of Technology, Pasadena, CA 91125, USA}

\author[0000-0001-8832-4488]{Daniel Huber}
\affiliation{Institute for Astronomy, University of Hawai`i, Honolulu, HI 96822, USA}

\author[0000-0002-0531-1073]{Howard Isaacson}
\affiliation{Department of Astronomy, University of California Berkeley, Berkeley, CA 94720, USA}
\affiliation{Centre for Astrophysics, University of Southern Queensland, Toowoomba, QLD, Australia}

\author[0000-0002-7084-0529]{Stephen R. Kane}
\affiliation{Department of Earth and Planetary Sciences, University of California, Riverside, CA 92521, USA}

\author{Molly Kosiarek}
\affiliation{Department of Astronomy and Astrophysics, University of California Santa Cruz, Santa Cruz, CA 95060, USA}

\author[0000-0001-8342-7736]{Jack Lubin}
\affiliation{Department of Physics \& Astronomy, University of California Irvine, Irvine, CA 92697, USA}

\author{Andrew Mayo}
\affiliation{Department of Astronomy, University of California Berkeley, Berkeley, CA 94720, USA}

\author[0000-0003-4603-556X]{Teo Mo\v{c}nik}
\affiliation{Gemini Observatory/NSF's NOIRLab, 670 N. A'ohoku Place, Hilo, HI 96720, USA}

\author[0000-0001-8898-8284]{Joseph M. Akana Murphy}
\altaffiliation{NSF Graduate Research Fellow}
\affiliation{Department of Astronomy and Astrophysics, University of California Santa Cruz, Santa Cruz, CA 95060, USA}

\author[0000-0001-9771-7953]{Daria Pidhorodetska}
\affiliation{Department of Earth and Planetary Sciences, University of California, Riverside, CA 92521, USA}

\author[0000-0001-7047-8681]{Alex S.\ Polanski}
\affiliation{Department of Physics \& Astronomy, University of Kansas, Lawrence, KS 66045, USA}

\author{Malena Rice}
\altaffiliation{NSF Graduate Research Fellow}
\affiliation{Department of Astronomy, Yale University, New Haven, CT 06520, USA}

\author[0000-0003-0149-9678]{Paul Robertson}
\affiliation{Department of Physics \& Astronomy, University of California Irvine, Irvine, CA 92697, USA}

\author[0000-0001-8391-5182]{Lee J.\ Rosenthal}
\affiliation{Department of Astronomy, California Institute of Technology, Pasadena, CA 91125, USA}

\author[0000-0001-8127-5775]{Arpita Roy}
\affiliation{Space Telescope Science Institute, Baltimore, MD 21218, USA}
\affiliation{Department of Physics and Astronomy, Johns Hopkins University, Baltimore, MD 21218, USA}

\author[0000-0003-3856-3143]{Ryan A. Rubenzahl}
\altaffiliation{NSF Graduate Research Fellow}
\affiliation{Department of Astronomy, California Institute of Technology, Pasadena, CA 91125, USA}

\author[0000-0003-3623-7280]{Nicholas Scarsdale}
\affiliation{Department of Astronomy and Astrophysics, University of California Santa Cruz, Santa Cruz, CA 95060, USA}

\author[0000-0002-1845-2617]{Emma V. Turtelboom}
\affiliation{Department of Astronomy, University of California Berkeley, Berkeley, CA 94720, USA}

\author[0000-0003-0298-4667]{Dakotah Tyler}
\affiliation{Department of Physics \& Astronomy, University of California Los Angeles, Los Angeles, CA 90095, USA}

\author[0000-0002-4290-6826]{Judah Van Zandt}
\affiliation{Department of Physics \& Astronomy, University of California Los Angeles, Los Angeles, CA 90095, USA}

\author[0000-0002-3725-3058]{Lauren M. Weiss}
\affiliation{Department of Physics and Astronomy, University of Notre Dame, Notre Dame, IN 46556, USA}

\author[0000-0001-7961-3907]{Samuel W.\ Yee}
\affiliation{Department of Astrophysical Sciences, 4 Ivy Lane, Princeton University, Princeton, NJ 08544 USA}

\begin{abstract}
We present the stellar and planetary properties for 85 TESS Objects of Interest (TOIs) hosting 108 planet candidates which comprise the TESS-Keck Survey (TKS) sample. We combine photometry, high-resolution spectroscopy, and \textit{Gaia} parallaxes to measure precise and accurate stellar properties. We then use these parameters as inputs to a lightcurve processing pipeline to recover planetary signals and homogeneously fit their transit properties. Among these transit fits, we detect significant transit-timing variations among at least three multi-planet systems (TOI-1136, TOI-1246, TOI-1339) and at least one single-planet system (TOI-1279). We also reduce the uncertainties on planet-to-star radius ratios $R_\textrm{p}/R_\star$ across our sample, from a median fractional uncertainty of 8.8$\%$ among the original TOI Catalog values to 3.0$\%$ among our updated results. With this improvement, we are able to recover the Radius Gap among small TKS planets and find that the topology of the Radius Gap among our sample is broadly consistent with that measured among \textit{Kepler} planets. The stellar and planetary properties presented here will facilitate follow-up investigations of both individual TOIs and broader trends in planet properties, system dynamics, and the evolution of planetary systems. 
\end{abstract}


\section{Introduction}
\label{sec:intro}

The NASA Transiting Exoplanet Survey Satellite (TESS; \citealt{Ricker15}) is currently in its 5$^\textrm{th}$ year of carrying out an all-sky survey in search of transiting planets orbiting nearby bright stars. So far, TESS has revealed over 6,000 planet candidates, building upon the legacies of its predecessors, NASA's \textit{Kepler} (\citealt{Borucki10}) and \textit{K2} (\citealt{Howell14}). These TESS Objects of Interest (TOIs) are now the subjects of numerous follow-up studies to verify their planetary nature and measure their properties (e.g. \citealt{Kane21}; \citealt{Teske21}; \citealt{Chontos22}; \citealt{Yee22}). The precise characterization of planet properties, however, is greatly dependent on the robustness of the input stellar parameters and lightcurve modeling procedure. 

The TESS Input Catalog (TIC; \citealt{Stassun18}, \citealt{Stassun19}; \citealt{https://doi.org/10.26134/exofop3}) contains stellar parameters for all TESS targets which were measured from broadband photometry and parallaxes. These stellar properties were used as inputs to the TESS data processing pipeline developed by the Science Processing Operations Center (SPOC; \citealt{Jenkins16}), which flags planet candidates as TOIs and performs an initial characterization (\citealt{Guerrero21}; \citealt{toi}). The propagation of large uncertainties throughout this process can lead to miscalculated or poorly constrained planet radii for TOIs, with a median fractional radius uncertainty of $\sigma(R_\textrm{p})/R_\textrm{p} \approx 7.4\%$ and a mean fractional uncertainty of $\sim$20.2$\%$. This large discrepancy between the median and mean uncertainty is due to a significant high-$\sigma(R_\textrm{p})$ tail, with $\sim$1 in 5 TOIs having $\sigma(R_\textrm{p})/R_\textrm{p} > 20\%$.

The TESS-Keck Survey (TKS) has set out to confirm and characterize a sample of 108 TOIs orbiting 85 TESS host stars (\citealt{Chontos22}). The TKS collaboration aims to precisely measure the stellar properties of this sample using spectra taken with the High-Resolution Spectrograph (HIRES; \citealt{Vogt94}) on the Keck I telescope at the W. M. Keck Observatory. These spectra also allow us to dynamically confirm the planetary nature of the TKS TOIs through the measurement of precise radial velocities (RVs), leading to numerous planet discoveries among the TKS team (e.g. \citealt{Dalba20}; \citealt{Dai20}; \citealt{Weiss21}; \citealt{Rubenzahl21}; \citealt{Scarsdale21}; \citealt{MacDougall21}; \citealt{Dalba22}; \citealt{Lubin22}; \citealt{Turtelboom22}; \citealt{MacDougall22}; \citealt{VanZandt22}).

To ensure consistency among the planetary and stellar parameters reported for TKS targets, we use the stellar parameters measured with our spectroscopic constraints as inputs to a lightcurve detrending and transit fitting pipeline applied homogeneously to all TKS targets. The planet properties derived by our pipeline allow us to begin addressing the major TKS Science Cases (see \citealt{Chontos22}), such as improving our understanding of planet bulk compositions, system architectures and dynamics, planetary atmospheres, and the role of stellar evolution in shaping planetary systems.

In this work, we describe our homogeneous characterization of both planetary and stellar properties of TKS targets, derived from transit photometry and single stellar spectra. In \S\ref{sec:stellar}, we update the stellar properties reported by \citealt{Chontos22} with additional photometric constraints and compare these results against those reported by the TOI Catalog and \textit{Gaia} Data Release 2. We also derive estimates of the quadratic limb darkening parameters for each TOI host to set priors on our transit model. In \S\ref{sec:lightcurves}, we describe the data processing pipeline used to retrieve, detrend, and fit the lightcurve photometry for our target sample. We account for both stellar variability and transit-timing variations in order to properly model the transit signals. We use the results from our transit fits and stellar characterization to derive various planet parameters in \S\ref{sec:planet}, including planet radius which we measure to a median uncertainty of 3.8$\%$. In \S\ref{sec:discussion}, we offer some preliminary insight gained through this analysis to begin addressing the science goals of TKS which will be evaluated in more depth in future works by the TKS team. Planet masses, precise eccentricities, and non-transiting companions are not addressed in this transit-focused study, and we leave that analysis to a full radial velocity analysis by TKS collaborators (Polanski et al. 2023, in prep).

\section{Stellar Properties}
\label{sec:stellar}

\subsection{Spectroscopic Constraints}
\label{sec:specmatch}

The TKS collaboration has collected spectra for the 85 TOIs from \cite{Chontos22} at a spectral resolution of $R = 50,000$ using the HIRES instrument at the W.M. Keck Observatory (\citealt{Vogt94}) from Summer 2019 to Fall 2022. We used iodine-free reconnaissance spectra, with S/N $\approx 40$/pixel across 3600 -- 9000 \AA, to check for rapid stellar rotation and rule out spectroscopic false-positives. We also extracted stellar effective temperature $T_\textrm{eff}$, metallicity [Fe/H], and surface gravity log$g$ from these spectra using two different methods according to the estimated effective temperature per target. The two methods used were \texttt{SpecMatch-Syn} (synthetic; \citealt{Petigura15}) and \texttt{SpecMatch-Emp} (empirical; \citealt{Yee17}). \texttt{SpecMatch-Syn} compares a given stellar spectrum to synthetic spectra generated by interpolating within a grid of modeled spectra from a library described by \cite{Coelho05}. \texttt{SpecMatch-Emp} fits a stellar spectrum via direct comparison to observed optical spectra from a dense library of stars with well-measured properties.

We initially processed the spectra of each target using both techniques and used their respective valid ranges of effective temperature to determine which set of derived properties is more reliable for a given star. Specifically, we use \texttt{SpecMatch-Syn} measurements for stars with estimated effective temperature between 4800 -- 6500 K, as determined by \texttt{SpecMatch-Emp}. For stars with estimated $T_\textrm{eff}$ beyond this range (i.e. $T_\textrm{eff}$ < 4800 K and $T_\textrm{eff}$ > 6500 K), we use \texttt{SpecMatch-Emp} measurements. In cases where \texttt{SpecMatch-Emp} is used, the \texttt{SpecMatch} model does not output an estimate of log$g$.

\subsection{Isochrone Modeling}
\label{sec:isochrone}

\subsubsection{Input values}
\label{sec:isoclassify}

We derive additional stellar properties such as mass $M_{\star}$, radius $R_{\star}$, density $\rho_{\star}$, luminosity $L_{\star}$, age, and extinction $A_v$ using \texttt{isoclassify} (\citealt{Berger20a}; \citealt{Huber17}). This software interpolates between the MESA Isochrones and Stellar Tracks models (MIST; \citealt{Dotter16}; \citealt{Choi16a}) to measure the best-fitting solution of stellar properties for a given set of input spectroscopic and photometric constraints. For M-dwarfs, \texttt{isoclassify} uses empirical relations from \cite{Mann19} to derive fundamental stellar properties (see \citealt{Berger20a} for details). \texttt{isoclassify} has been shown to produce robust results and was the foundation for deriving the \emph{Gaia}-\emph{Kepler} stellar properties catalog (\citealt{Berger20a}). 

We used the outputs of the preferred \texttt{SpecMatch} model per target as priors to constrain the isochrone grid space explored by our \texttt{isoclassify} models. This includes priors on $T_\textrm{eff}$ and [Fe/H] from either \texttt{SpecMatch} method, along with a prior on log$g$ when \texttt{SpecMatch-Syn} is preferred. We also include priors on stellar parallax drawn from \textit{Gaia} DR2, with a median uncertainty of 0.03 mas. We incorporate photometric inputs for several all-sky photometric bands from the \textit{2MASS} (\citealt{Skrutskie06}) and \textit{Gaia} missions to further constrain our \texttt{isoclassify} results. These bands are \textit{2MASS} $J$, $H$, $K$ and \textit{Gaia} DR2 $G$, $B_\textrm{p}$, $R_\textrm{p}$, from which various photometric colors are used to calibrate the underlying isochrone model grid. We also select \textit{2MASS} $K$ as the photometric band to be used for absolute magnitude calculations during \texttt{isoclassify} modeling. All input parameter priors are assumed to be normally distributed.

\begin{figure}[ht]
\centering
\includegraphics[width=0.47\textwidth]{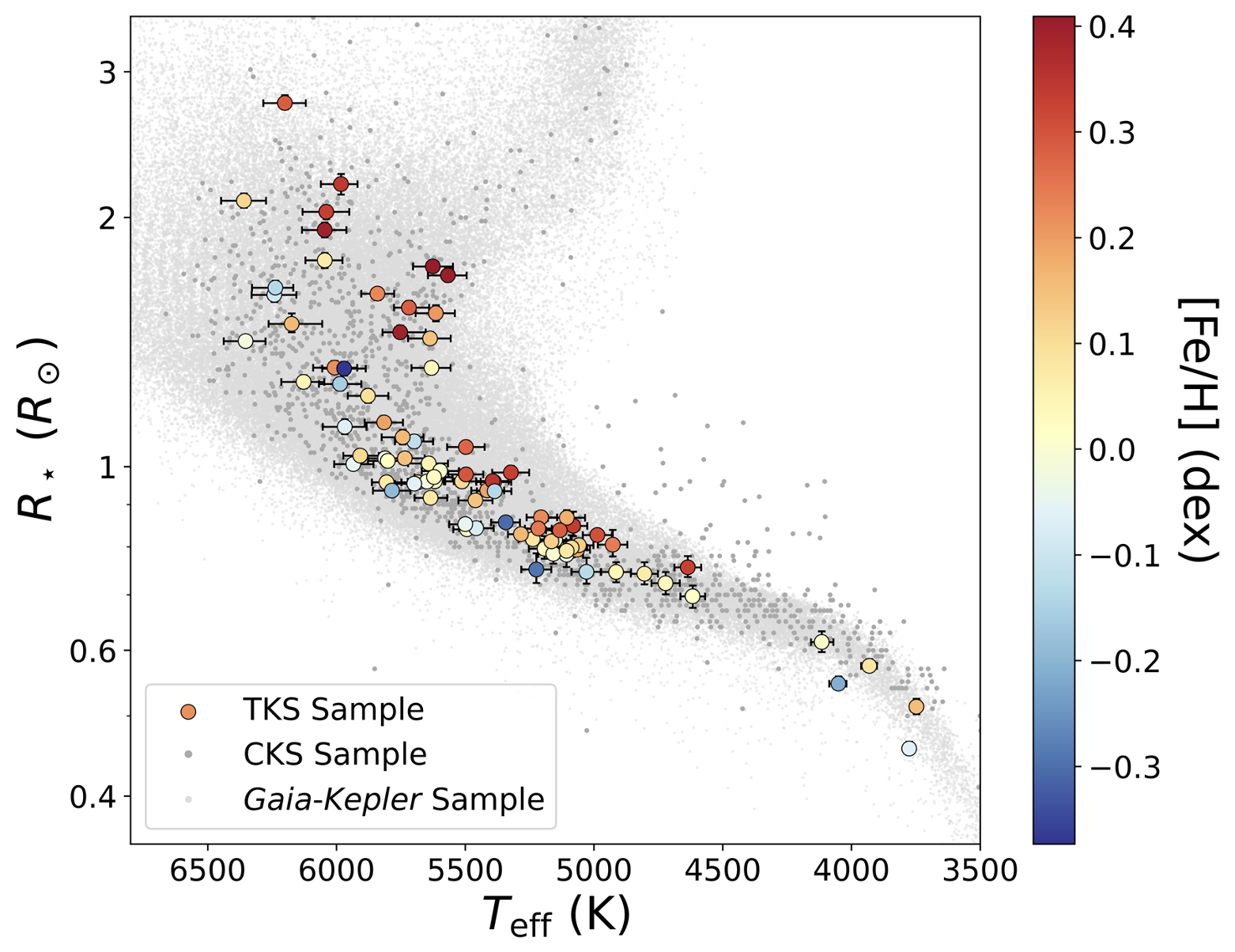}
\caption{Stellar radius as a function of effective temperature for the 85 host stars in the TKS sample, with metallicity shown as color gradient. The California-\emph{Kepler} Survey sample (\citealt{Petigura22}) and the \emph{Gaia}-\emph{Kepler} sample (\citealt{Berger20}) are shown for reference in dark and light grey, respectively.}
\label{fig:rad-teff-feh}
\end{figure}

The inclusion of multiple photometric constraints allows us to fit for extinction directly using \texttt{isoclassify} without needing to specify a generalized dust map (see, e.g. \citealt{Lallement19}), making our extinction and metallicity estimates more accurate on a target-by-target basis. This improves the accuracy of our stellar characterizations as compared to \cite{Chontos22} where only a single photometric band was used with a generalized dust map. The median uncertainty for the \textit{2MASS} photometry that we use is 0.02 mag. To account for systematic zero-point differences in photometric systems, we use a standardized uncertainty floor of 0.01 mag for \textit{Gaia} photometry.

\begin{figure*}[ht]
\centering
\includegraphics[width=0.95\textwidth]{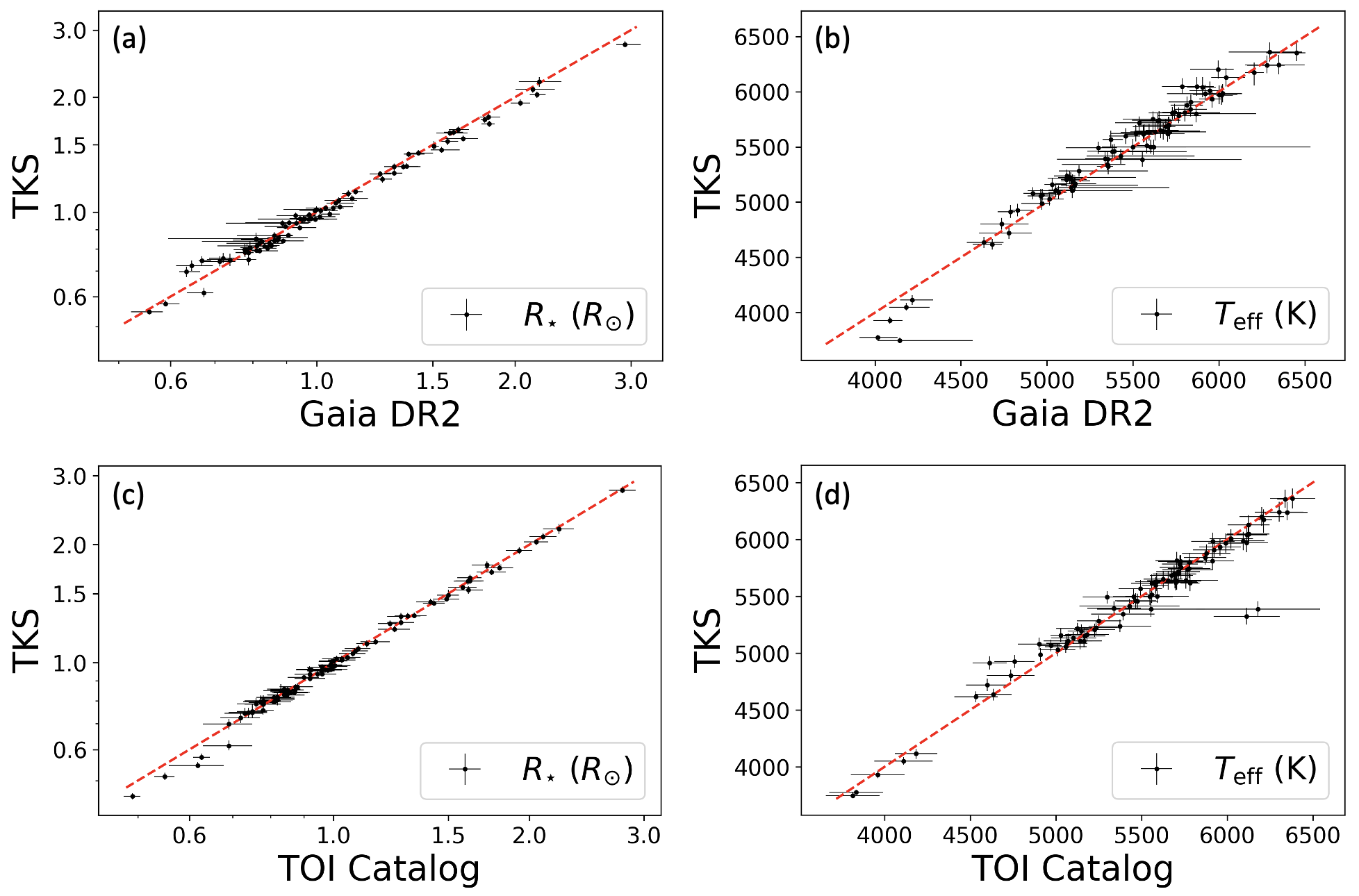}
\caption{Stellar radius (left) and effective temperature (right) as measured by our stellar characterization procedure (TKS) on the y-axis compared to the values reported by the \textit{Gaia} DR2 pipeline (top) and the TOI Catalog (bottom) on the x-axis. The TKS results produce reduced uncertainties and are generally consistent with one or both external sources. The residuals of our comparison between $R_{\star, \textrm{TKS}}$ and $R_{\star, \textrm{TOI}}$ can be seen in Figure \ref{fig:residuals}a. We also note that the \textit{Gaia} DR2 pipeline does not report stellar properties when $R_{\star, Gaia} \lesssim 0.5 R_\odot$, leaving out two TKS targets.}
\label{fig:rad-teff}
\end{figure*}

With these inputs to \texttt{isoclassify}, we perform stellar characterization via isochrone grid modeling to measure $M_{\star}$, $R_{\star}$, $\rho_{\star}$, $L_{\star}$, age, and $A_v$ for all 85 TKS targets, in addition to spectroscopically constrained $T_\textrm{eff}$, [Fe/H], and log$g$. This catalog of precise homogeneously derived stellar properties will serve as a reference for all future TKS studies and as a resource for the broader astronomy community. We show the relationship between $R_{\star}$--$T_\textrm{eff}$--[Fe/H] in Figure \ref{fig:rad-teff-feh} for this sample, as compared to the distributions of both the California-\emph{Kepler} Survey sample (\citealt{Petigura17}) and the \emph{Gaia}-\emph{Kepler} sample (\citealt{Berger20}). We note that the TKS sample includes a broad range of stellar types, similar to the distributions of stars studied of similar past works.

\subsubsection{Uncertainties}
\label{sec:uncertainties}

The median uncertainties that we measure for $R_{\star}$, $M_{\star}$, $T_\textrm{eff}$, and log$g$ are reported in Table \ref{tab:uncertainties}, as compared against the original TOI Catalog source uncertainties (\citealt{Guerrero21}; \citealt{toi}). We do not currently account for model-dependent uncertainties associated with the MIST models to maintain consistency with the reported uncertainties by the TOI Catalog and \textit{Gaia} DR2 catalog. The appropriate corrections (described by \citealt{Tayar20}) can be added in quadrature to our error values to encapsulate this systematic error. When we perform this correction, we find updated median fractional uncertainties of 4.5$\%$, 4.0$\%$, 1.3$\%$, and 0.6$\%$ for $R_{\star}$, $M_{\star}$, $T_\textrm{eff}$, and log$g$, respectively. Most notable among these is the change in the stellar radius uncertainty, which increased from 1.7$\%$ to 4.5$\%$ when the systematic error was introduced. This ultimately propagates to an increase in the median $R_\textrm{p}$ uncertainty as well, from 3.8$\%$ to 5.8$\%$. Even with this additional systematic uncertainty on $R_\textrm{p}$, our results remain better constrained compared to the TOI Catalog.

We also note that two targets from the TKS sample have missing or erroneous \textit{2MASS} catalog uncertainties. TOI-1807 has an anomalously large $K$-mag uncertainty ($\sigma_\textrm{K}$ = 10 mag) and TOI-1246 is missing an uncertainty value for $J$-mag. We replaced both of these with a standardized uncertainty value of 0.02 mag, roughly consistent with the median errors for these bands among our sample.

\subsubsection{Comparison with TOI Catalog and \textit{Gaia} results}
\label{sec:comparison}

To validate our stellar parameters, we compared our final values of $T_\textrm{eff}$ and $R_{\star}$ from \texttt{isoclassify} with the TOI Catalog and \textit{Gaia} DR2. We show these comparisons in Figure \ref{fig:rad-teff}. The stellar parameters measured from these external sources were calculated using either photometry with parallaxes (TOI Catalog; \citealt{Stassun18}) or a combination of photometry with parallaxes and low-resolution spectra (\textit{Gaia} DR2; \citealt{BailerJones13}). 

For each comparison, we calculated the ratio of the given parameter between our TKS results and the external source data (i.e. $R_{\star, \textrm{TKS}}/R_{\star, \textrm{TOI}}$) to demonstrate the lack of bias in these results. We measured the mean and standard deviation of these ratio distributions for each comparison and found: (Fig. \ref{fig:rad-teff}a) $R_{\star, \textrm{TKS}}/R_{\star, Gaia} = 0.99 \pm 0.04$, (Fig. \ref{fig:rad-teff}b) $T_{\textrm{eff}, \textrm{TKS}}/T_{\textrm{eff}, Gaia} = 1.00 \pm 0.01$, (Fig. \ref{fig:rad-teff}c) $R_{\star, \textrm{TKS}}/R_{\star, \textrm{TOI}} = 0.99 \pm 0.03$, and (Fig. \ref{fig:rad-teff}d) $T_{\textrm{eff}, \textrm{TKS}}/T_{\textrm{eff}, \textrm{TOI}} = 1.00 \pm 0.02$. The scatter between our TKS results and the external data source for each comparison was consistent with the reported parameter uncertainty. This implies a general agreement between our TKS stellar parameters and those reported by external data sources. We do find that the \textit{Gaia} DR2 pipeline produces overestimated $T_\textrm{eff}$ for cooler TKS stars, but this is consistent with a known degradation in the precision of the \textit{Gaia} DR2 pipeline around $T_\textrm{eff} \approx 4000$ K (\citealt{Andrae18}). 

From these comparisons, we flag two targets with discrepancies of >3$\sigma$ between our results and the values reported from the external data sources. TOI-2145 had an estimated effective temperature whose uncertainty spanned across the threshold of reliability between \texttt{SpecMatch-Emp} and \texttt{SpecMatch-Syn}. Selecting between these two methods following the criteria in \S\ref{sec:specmatch} yielded a radius measurement for this target that was inconsistent with both the TOI Catalog and \textit{Gaia} DR2 estimates. By manually swapping methods and allowing for an exception to these rules, we found good alignment between our final radius measurement and the values reported from the other two sources. The remaining flagged target (TOI-2114), which can be seen in Figure \ref{fig:rad-teff}d, has a discrepant $T_\textrm{eff}$ when we compare our result to the TOI Catalog. However, our result remains highly consistent with the $T_\textrm{eff}$ reported by \textit{Gaia} DR2 and other external sources. Given this observation and the relatively large uncertainties reported by the TOI Catalog, we attribute the disagreement to slight differences in modeling methods and proceed without making any adjustments for this particular target.

\subsection{Limb Darkening}
\label{sec:limbdarkening}

Stellar limb darkening is the decrease in brightness of the disk of star from its center towards its edge or limb. A set of quadratic coefficients is often used to define this limb darkening function, which describes the ingress and egress of a transit profile (\citealt{Mandel02}). An accurate estimate of limb darkening coefficients is crucial for measuring accurate planet properties, especially for planets that transit closer to the limb of a star (e.g. higher impact parameter or inclination).

There are two leading stellar atmosphere models that are widely used for calculating quadratic limb darkening coefficients: the PHOENIX model (\citealt{HauschildtBaron99}) as implemented by \cite{Husser13} and the ATLAS models (\citealt{CastelliKurucz03}) as implemented by \cite{ClaretBloemen11}. Investigations into the accuracy of these two methods (see, e.g., \citealt{PatelEspinoza22}) have demonstrated that such models produce quadratic limb darkening coefficients with an average discrepancy of $\sim$0.1. For reference, a typical quadratic limb darkening value for a solar-like star is in the range 0.1 -- 0.5, so an uncertainty of 0.1 is typically >20$\%$ of the nominal measurement. Additionally, this formal uncertainty between methods only encompasses part of the true uncertainty that arises when measuring these coefficients from atmospheric models. With this in mind, we use the \texttt{limb-darkening} code (\citealt{EspinozaJordan15}) to easily compute both the PHOENIX-derived and ATLAS-derived limb darkening coefficients for all targets in our sample. This derivation requires several stellar inputs, which we draw from our \texttt{isoclassify} results: $T_\textrm{eff}$, [Fe/H], and log$g$. We then use these values to adopt nominal estimates of coefficients \{$u_1$, $u_2$\} for each target with a systematic noise floor of 0.1.

\begin{deluxetable}{llcc}
\tabletypesize{\scriptsize}
\tablecaption{Median Fractional Uncertainties\label{tab:uncertainties}}
\tablehead{\colhead{} & \colhead{Parameter} & \colhead{TKS Result Unc.} & \colhead{TOI Catalog Unc.} \\ \colhead{} & \colhead{} & \colhead{\%} & \colhead{\%}}
\startdata
$Star$ &     $R_\star$ & 1.7 & 5.1 \\
  &     $M_\star$ & 3.0 & 12.6 \\
  & $T_\textrm{eff}$ & 1.2 &  2.3 \\
  &        log$g$ & 0.52 &  1.9 \\
\hline
$Planet$  &         $R_\textrm{p}$ & 3.8 &  6.7 \\
  &  $R_\textrm{p}/R_\star$ & 3.0 &  8.8 \\
  &  $T_{14}$ & 1.8 &  9.7 \\
  &           $P$ & $2.2\times10^{-4}$ & $3.8\times10^{-4}$ \\
  &         $t_0$ & $7.1\times10^{-5}$ &  $7.3\times10^{-5}$ \\
\enddata
\tablecomments{Reported fractional uncertainties for several stellar and planetary parameters compared between our TKS results and those reported in the TOI Catalog, demonstrating our improved precision across the TKS sample. The values presented here do not account for systematic uncertainties that may arise from differences in underlying stellar model grids (see \S\ref{sec:uncertainties}).}
\end{deluxetable}

Given the large uncertainty in these measurements, we establish a simple set of rules to determine which set of limb darkening coefficients ($u_i$) to use in deriving our final catalog of system properties. (1) When |$u_{i, \textrm{ATLAS}}$ - $u_{i, \textrm{PHOENIX}}$| < 0.1, we select AVG($u_{i, \textrm{ATLAS}}$, $u_{i, \textrm{PHOENIX}}$) as the final coefficient values. (2) When |$u_{i, \textrm{ATLAS}}$ - $u_{i, \textrm{PHOENIX}}$| $\geq$ 0.1, we select $u_{i, \textrm{ATLAS}}$ as the final coefficient values. These rules were established upon consideration of the systematic comparison of these two methods carried out by \citealt{PatelEspinoza22}. We also assume a standardized uncertainty of 0.1 for all limb darkening coefficients among the TKS sample, which is later propagated into a model prior during our transit fitting procedure.

\begin{figure*}[ht]
\centering
\includegraphics[width=0.9\textwidth]{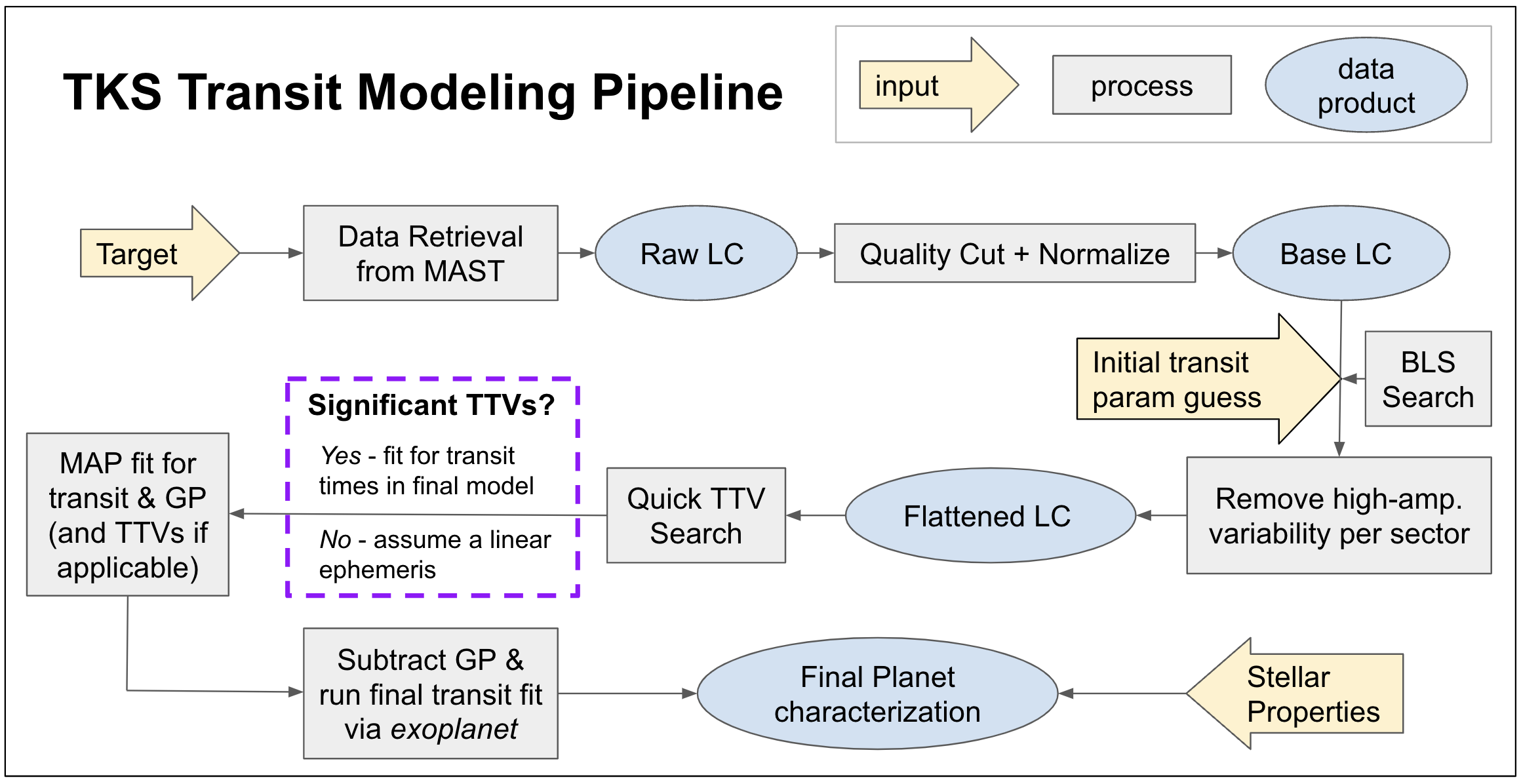}
\caption{Diagram demonstrating the flow of data throughout the TKS light curve modeling pipeline. Yellow arrows are data inputs, grey boxes are processes, and blue ovals are data products.}
\label{fig:pipeline}
\end{figure*}

\section{Light Curve Modeling}
\label{sec:lightcurves}

\subsection{Input Data}
\label{sec:photometry}

We perform our light curve detrending and modeling using transit photometry from Sectors 1--60 of the TESS mission. We accessed the Pre-search Data Conditioning Simple Aperture Photometry (PDC-SAP; \citealt{Stumpe12}; \citealt{Stumpe14}; \citealt{Smith12}) through the Mikulski Archive for Space Telescopes (MAST). For the majority of TKS targets, we used 2-min-cadence time-series photometry processed by the primary TESS Science Processing Operations Center pipeline (SPOC; \citealt{Jenkins16}). For TOI-1386, 1601, and 2045, photometry from the primary SPOC pipeline was unavailable, so we instead used PDC-SAP photometry from the TESS-SPOC updated pipeline which produces lightcurves for targets selected from full-frame image observations (\citealt{Caldwell20}).

\subsection{False-Positive Vetting}
\label{sec:fp-vetting}

While false-positive vetting is not a primary focus of this work, we acknowledge that the resolved false-positive rate amongst \textit{all} TOIs is on the order of $\sim$20--30$\%$ (\citealt{toi}; \citealt{Cacciapuoti22}; \citealt{Magliano23}). However, this is an extreme upper bound for the TKS sample. We specifically selected our sample to include \textit{only} TOIs with unambiguous planetary dispositions confirmed by both the initial TOI vetting pipeline (\citealt{Guerrero21}) and the TESS Follow-up Program Working Group.

Additionally, all TKS targets were individually vetted by members of the TKS collaboration, as detailed in \cite{Chontos22}. This vetting process allowed us to establish a set of selection rules to minimize the risk of including unresolved false-positives in our final target sample by making cuts based on transit S/N, out-of-transit centroid offsets, and the proximity of stellar companions. Therefore, we suspect that the false-positive rate in the TKS sample is much lower than the quoted upper bound, but we leave a quantitative assessment of this value for later work. Should any of the TKS TOIs be deemed false-positives, the planetary properties presented here would need to be revisited.

\subsection{Lightcurve Pre-Processing}
\label{sec:preprocessing}

We apply the following pre-processing routine homogeneously across all lightcurves for the 85 TKS targets. A complete outline of this pre-processing routine and our subsequent modeling pipeline can be found in a flowchart in Figure \ref{fig:pipeline}. We first mask all points that were flagged for poor quality (quality flag > 0), with the exception of TOI-1456 for which a valid transit signal is mistakenly flagged as scattered light contamination by the SPOC pipeline (see \citealt{Dalba20}). We then normalize each lightcurve sector iteratively and subtract 1 from the normalized flux values such that $\mu_\textrm{flux} = 0$ for all sectors. We also flag four lightcurves as having high stellar variability, which we determine based on whether or not the standard deviation of a target's out-of-transit flux is $>$3$\sigma_\textrm{flux}$, where $\sigma_\textrm{flux}$ is the median uncertainty of the individual photometric data points. These four targets are TOI-1136, 1726, 1807, and 2076. We also apply an upper sigma cut of 5$\sigma$ to remove high photometric outliers among targets that do not meet this high-variability criteria.

For light curves with low stellar variability, we perform a box least squares (BLS; \citealt{Kovacs02}) search to determine initial estimates of the orbital period and first transit mid-point of each planet. We compare the BLS search results to estimates taken from the primary TOI Catalog (\citealt{toi}), flagging any results with >3$\sigma$ discrepancy when compared to this source. We then follow up flagged targets with a manual inspection, determining that all discrepancies are likely the result of new photometry assisting in pinning down the orbital period with greater precision. In cases of high stellar variability, we do not perform this initial BLS search and instead proceed with the values drawn directly from the TOI Catalog or other literature as our initial guesses.

We use these period and mid-point values, along with transit durations from the TOI catalog, to mask out all transits with a mask width of $\pm0.75$ times the transit duration from the transit mid-point. While this mask does not yet account for transit-timing variations (TTVs), we use it here for an initial detrending step which we improve upon in our subsequent fit in \S\ref{sec:fullmodel}. With the masked photometry, we modeled stellar variability in each lightcurve sector using a Gaussian Process (GP) via \texttt{celerite2} (\citealt{celerite2}) through the \texttt{exoplanet} interface (\citealt{Foreman-Mackey2021}). The GP fit includes a term to model the variability via a stochastically-driven damped harmonic oscillator with quality factor $Q = 1/\sqrt{2}$, with log-normal priors on both the undamped period ($\rho_\textrm{GP}$) and the standard deviation of the process ($\sigma_\textrm{GP}$) with means $\mu_{\rho_\textrm{GP}} = 0$ and $\mu_{\sigma_\textrm{GP}} =$ ln($\sigma_\textrm{flux}$), respectively. We interpolate over the masked transits to create a smooth fit to the variability for each lightcurve sector and subtract this trend from the original unmasked data to produce an initial flattened lightcurve. We stitch together the flattened photometry for each sector and save the maximum \emph{a posteriori} (MAP) parameters for the GP fit which we use later as informed initial guesses in a more complete lightcurve model.

\begin{deluxetable}{ll}
\tabletypesize{\scriptsize}
\tablecaption{Transit Model Priors\label{tab:priorstab}}
\tablehead{\colhead{Parameter}\hspace{1cm} & \colhead{Prior}}
\startdata
$P$ (days) & log$P$ $\sim$ N(log$P_\textrm{init}$, 1) \\
$t_0$ (days) & $t_0$ $\sim$ N($t_\textrm{0, init}$, 1) \\
$R_\textrm{p}/R_{\star}$ & log($R_\textrm{p}/R_{\star}$) $\sim$ U(-9, 0) \\
$b$ & $b$ $\sim$ U(0, $1+R_\textrm{p}/R_{\star}$) \\
$T_{14}$ (days) & log$T_{14}$ $\sim$ U(-9, 0) \\
$u_{i}$ & $u_{i}$ $\sim$ N($u_{i, \textrm{init}}$, 0.1) \\
$\mu$ & $\mu$ $\sim$ N(0, 1) \\
$\sigma_\textrm{LC}$ & log$\sigma_\textrm{LC}$ $\sim$ N(log$\sigma_\textrm{flux}$, 1.0) \\
\enddata
\tablecomments{Priors on final transit model parameters. $P$, $T_{14}$, and $t_0$ are all given in units of days, with $t_0$ using the reference frame BJD - 2457000. Priors include normal (N) and uniform (U) distributions.}
\end{deluxetable}

\subsection{Estimating TTVs}
\label{sec:ttvs}

Along with determining an initial fit to the stellar variability trend, we also use the \texttt{exoplanet} lightcurve modeling package to find an initial MAP fit to the transit times for each planet. We use our previous $P$ and $t_0$ estimates to calculate a set of initial guesses for the transit times assuming a linear ephemeris. We model the true transit times for each planet using an initial Gaussian prior for each transit time, centered on their linear ephemeris values with a standard deviation of 0.1 day. Along with the transit times, we also fit the planet-to-star radius ratio ($R_\textrm{p}/R_\star$), impact parameter ($b$), transit duration from 1st-to-4th contact ($T_{14}$), quadratic limb darkening parameters ($u_1, u_2$), mean out-of-transit flux ($\mu$), and lightcurve jitter ($\sigma_\textrm{LC}$). The orbital period ($P$) and initial transit mid-point ($t_0$) are also measured as derived values. The priors used on all modeled parameters are given in Table \ref{tab:priorstab} and described in more detail in \S\ref{sec:fullmodel}. 

This MAP fit to the transit shape and transit times allows us to get a quick estimate of any TTVs detected for each planet. Based on these estimates, we determine for each planet whether to proceed with a final model that assumes a linear ephemeris or includes transit-time fits with a Gaussian prior. We use the following criteria to determine if the estimated TTVs warrant further consideration in our full lightcurve model: (1) more than 3 high-quality transits, (2) a TTV standard deviation greater than the photometric exposure time (i.e. 2 minutes), and (3) a single-transit signal-to-noise ratio of 2 or greater. Since we expect a linear ephemeris for most planets in our sample, we do not offer an in-depth analysis of these variations here and instead leave that analysis to later work (see \S\ref{sec:multis} for more details).

\begin{figure*}[ht]
\centering
\includegraphics[width=0.95\textwidth]{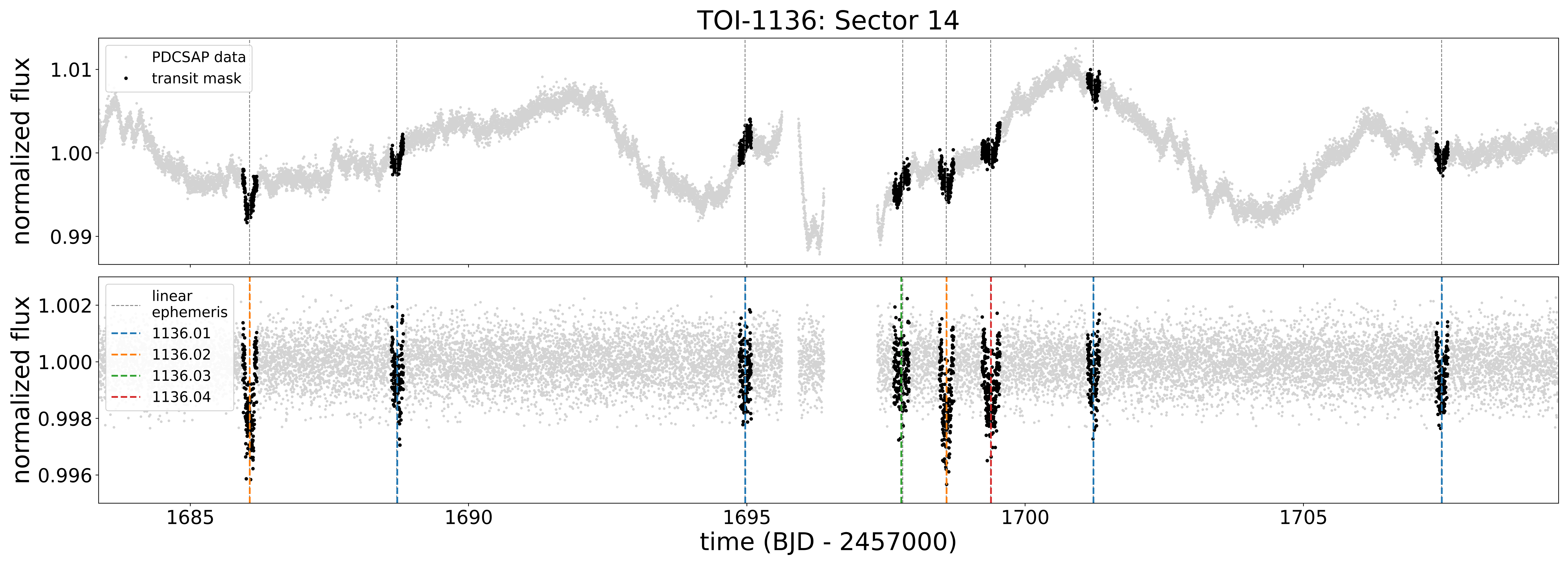}
\caption{Before (top) and after (bottom) applying our de-trending and initial TTV fitting procedure to the photometry of TOI-1136, a multi-planet system with high stellar variability. We only show the first sector (Sector 14) of this target's photometry here, but all sectors display similar variability. We model our de-trended transit photometry and measure $R_\textrm{p}/R_\star$ values for all 4 planets which agree with the values reported by \citealt{Dai22} within $\sim$1.5$\sigma$, as opposed to the TOI Catalog values which are all more than 1.5$\sigma$ discrepant. Black points highlight transits ($\pm$0.75 times $T_{14}$), colored vertical dashed lines show modeled transit mid-points, and dark grey vertical dashed lines show calculated transit mid-points assuming a linear ephemeris. In the sector shown, we find the most significant TTVs for TOI-1136.03 (green) as can be seen by the separation between the modeled and calculated mid-points.}
\label{fig:detrending}
\end{figure*}

\subsection{Full Lightcurve Model}
\label{sec:fullmodel}

We use the MAP values of all model parameters from both our trend model and TTV model as well-informed initial guesses for a complete MAP fit to the normalized lightcurve photometry. We simultaneously fit for the stellar variability trend, transit times, and transit shape using \texttt{exoplanet} with the same model priors as were used in our previous MAP fits. This full model allows us to measure photometric variability via Gaussian Processes (as was done in \S\ref{sec:preprocessing}) while also accounting for the true transit shapes and times. We subtract this MAP fit of the photometric variability to produce the final de-trended lightcurve for our analysis (see Figure \ref{fig:detrending} for an example). We create a transit mask using the MAP values from this full model fit to exclude out-of-transit photometry beyond $\pm2.5$ $T_{14, \textrm{MAP}}$ from each transit mid-point time, removing data that is uninformative of the transit fit in order to improve posterior sampling efficiency.

With our fully de-trended photometry and TTV estimates from this full MAP fit, we build our final model intended for posterior sampling and robustly fitting the transit shape. For planets with TTV estimates below the thresholds described in \S\ref{sec:ttvs}, we assume a linear ephemeris and do not directly fit transit times in the final model. Altogether, we fit for \{$P$, $t_0$, $R_\textrm{p}/R_\star$, $b$, $T_{14}$, $u_1, u_2$, $\mu$, $\sigma_\textrm{LC}$\} and sometimes transit times $\{TT_{i}\}$, conditioned on our fully de-trended TESS lightcurve.

We apply priors to each of the lightcurve model parameters, briefly outlined below and in Table \ref{tab:priorstab}. We apply Gaussian priors on $t_0$, $\mu$, $u_1$, and $u_2$, each centered on estimates that were reasonably well constrained from earlier analysis. We use a log-normal prior on $\sigma_\textrm{LC}$ to ensure positive values, with $\mu(\sigma_\textrm{LC}) =$ log($\sigma_\textrm{flux}$) and $\sigma(\sigma_\textrm{LC}) = 1$. For $R_\textrm{p}/R_\star$ and $T_{14}$, we use log-uniform priors with upper and lower bounds described in Table \ref{tab:priorstab}, designed to minimize the effect that these boundaries have on the completeness of the posterior sampling. We also model $b$ with a uniform prior from 0 to $1 + R_\textrm{p}/R_\star$ to encompass the regime of grazing transits in our posterior space.

Here, we use $T_{14}$ to directly fit the transit duration rather than measuring it through indirect means such as sampling in "circular" stellar density (see, e.g. \citealt{Dawson12}) or simultaneously fitting the true stellar density ($\rho_\star$), eccentricity ($e$), and argument of periastron ($\omega$). While the former method has been found to produce biased results (see \citealt{Gilbert22}), the latter is unbiased but requires three parameters to measure that which can be described by one ($T_{14}$) more efficiently. Our duration-sampling method also allows us to calculate posterior distributions for $e$ and $\omega$ post-modeling through importance sampling, described further in \S\ref{sec:planet}. Regardless of the method used, we emphasize that most \{$e$, $\omega$\} constraints measured from photometry alone are imprecise for individual planets and are better applied towards population-level studies (\S\ref{sec:ecc}). 

We sample the posterior probability density of all model parameters with \texttt{exoplanet} using the gradient-based No U-Turn Sampling method (\citealt{Hoffman11}; \citealt{Betancourt16}) as implemented by \texttt{PyMC3} (\citealt{pymc16}). For all posterior sampling performed in this work, we use 5,000 tuning steps with an additional 3,000 sampler draws and a target acceptance fraction of 0.95. This sampling process is performed via two sampler chains across two CPU cores, and we consider the process to be converged when the Gelman-Rubin statistic $\hat{R}$ for these chains is $\hat{R} < 1.01$ (\citealt{Gelman92}).

\begin{figure}[ht]
\centering
\includegraphics[width=0.47\textwidth]{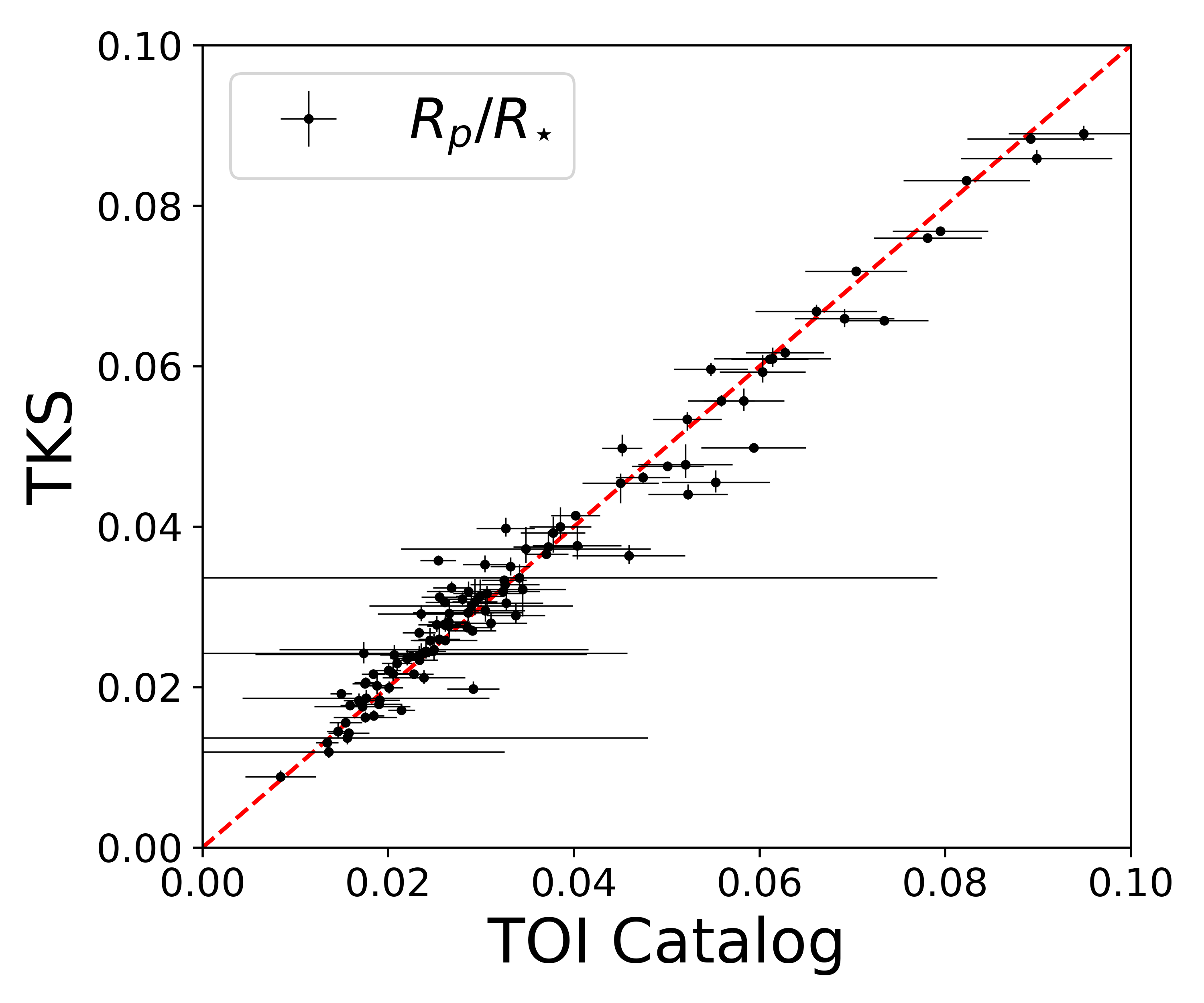}
\caption{Planet-to-star radius ratio $R_\textrm{p}/R_\star$ as measured by our planet characterization pipeline (TKS; y-axis) compared to the radius ratio values reported by the TOI Catalog (x-axis). The TKS results have significantly reduced uncertainties and are generally consistent with the TOI Catalog results (see \S\ref{sec:rprs}).} 
\label{fig:rprs-comp}
\end{figure}

\section{Planet Properties}
\label{sec:planet}

\subsection{Importance-Weighted Posterior Distributions}
\label{sec:impsamp}

Our final planet property measurements are based on the posterior distributions measured with \texttt{exoplanet}. These properties are modeled via a fit to the transit shape, conditioned on the input lightcurve data. However, our selected modeling basis does not inherently assume that any information is known about $\rho_\star$, $e$, or $\omega$ and, therefore, the posterior chains produced directly by our sampler can be considered \textit{unweighted}. Given our robust measurements of $\rho_\star$ from our stellar characterization, we can apply an "importance weight" (\citealt{OhBerger93}; \citealt{Gilks95}) to these posterior samples according to how well the \textit{unweighted} results can be described by the $\rho_\star$ values measured with \texttt{isoclassify}. We will refer to these precise stellar density measurements as $\rho_{\star, \textrm{iso}}$. The importance weights also allow us to produce properly weighted posterior distributions for $e$ and $\omega$ without needing to directly sample them during the lightcurve modeling -- a step that takes seconds rather than hours. This method has been used extensively in past literature for similar applications of improving modeling efficiency (see, e.g., \citealt{Ford05}; \citealt{Ford06}) or measuring eccentricity (see, e.g., \citealt{Dawson12}; \citealt{VE19}).

To determine the appropriate importance weights, we first calculate the estimated stellar density at each sampler step based on the sampled quantities $P$, $R_\textrm{p}/R_\star$, $b$, and $T_{14}$. We refer to this derived density as $\rho_{\star, \textrm{derived}}$. We calculate these values according to Equation \ref{eq:rhostar}, which is a rearrangement of the transit duration equation described by \cite{Winn10}:
\begin{equation}
\label{eq:rhostar}
\rho_{\star, \textrm{derived}} = \frac{3 \pi}{G P^2}\left(\frac{\left(1+R_\textrm{p}/R_\star\right)^2-b^2}{\sin^2{\left(\frac{T_{14} \pi}{P}\frac{1+e\sin{\omega}}{\sqrt{1-e^2}}\right)}}+b^2\right)^{3/2}.
\end{equation}
This equation, however, also includes $e$ and $\omega$, for which we do not yet have any information. Here, we draw random samples of $\{e, \omega\}$ from uniform prior distributions $e \sim U(0,1)$ and $\omega \sim U(-\frac{\pi}{2},\frac{3\pi}{2})$ to be used in the calculation. We prefer to use uninformative uniform priors on these parameters to remain agnostic to eccentricity here but will explore the parameterization of this eccentricity prior more in depth in MacDougall et al. 2023 (in prep). The stellar density values that we calculate here produce a derived posterior distribution based on our \textit{unweighted} posterior samples and uniform \{$e$, $\omega$\} samples.

We then compare the samples of $\rho_{\star, \textrm{derived}}$ against the independently measured $\rho_{\star, \textrm{iso}}$ for a given target by computing the log-likelihood of each $i^\textrm{th}$ sample,

\begin{equation}
\label{eq:log-like}
    \log \mathcal{L}_i = -\frac{1}{2}\Big(\frac{\rho_{\star, \textrm{derived}, i} - \rho_{\star, \textrm{iso}}}{\sigma(\rho_{\star, \textrm{iso}})}\Big)^2,
\end{equation}
assuming a Gaussian likelihood function. We then weight each sample from our original \textit{unweighted} posterior distributions by
\begin{equation}
\label{eq:weights}
    w_i = \frac{\mathcal{L}_i}{\sum_i \mathcal{L}_i}
\end{equation}
to produce the final importance-weighted posterior distributions for each parameter. We also apply these same weights to the random, uniform \{$e$, $\omega$\} samples to derive their weighted posterior distributions. All summary statistics that are reported or used throughout the remaining analysis are based on the 15$^\textrm{th}$, 50$^\textrm{th}$, and 85$^\textrm{th}$ percentiles of these importance-weighted posteriors.

\subsection{Planet-to-star Radius Ratio}
\label{sec:rprs}

\begin{figure*}[ht]
\centering
\includegraphics[width=0.95\textwidth]{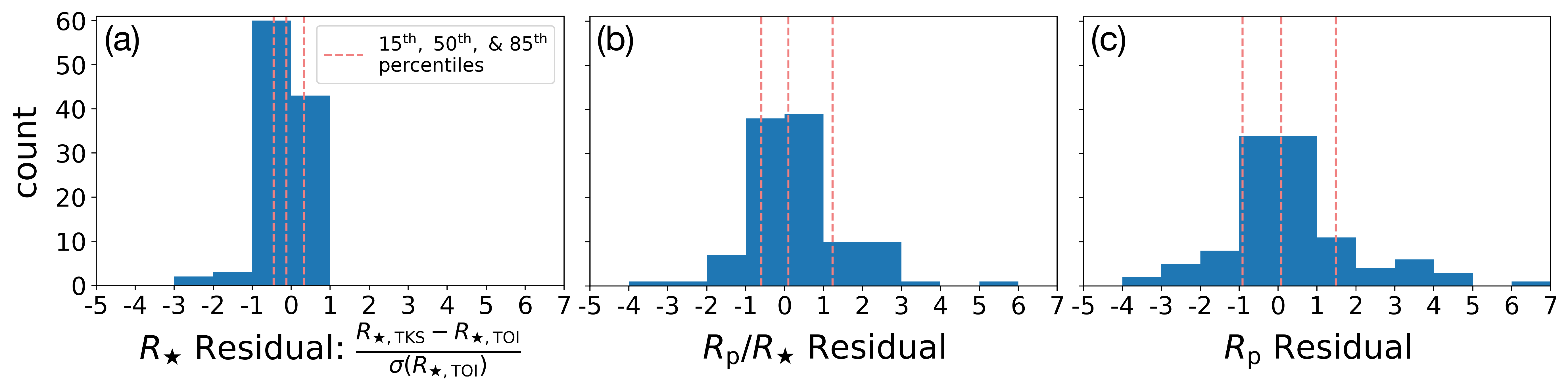}
\caption{Residuals of (a) stellar radius $R_\star$, (b) planet-to-star radius ratio $R_\textrm{p}/R_\star$, and (c) planet radius  $R_\textrm{p}$ between our results and the TOI Catalog measurements (i.e. residual = $(R_{\star, \textrm{TKS}} - R_{\star, \textrm{TOI}})/\sigma(R_{\star, \textrm{TOI}})$). Dashed vertical lines show the $15^\textrm{th}$, $50^\textrm{th}$, and $85^\textrm{th}$ percentiles of each distribution of residuals.}
\label{fig:residuals}
\end{figure*}

Before deriving final planet radii, we first confirm that our final modeled $R_\textrm{p}/R_\star$ values are unbiased and generally consistent with those reported by the TOI Catalog (Figure \ref{fig:rprs-comp}). We find no evidence of systematic bias in our results, measuring the average ratio between the TKS results and TOI Catalog values $(R_\textrm{p}/R_\star)_\textrm{TKS} / (R_\textrm{p}/R_\star)_\textrm{TOI} = 1.02 \pm 0.11$. This is consistent within the average fractional uncertainty of $R_\textrm{p}/R_\star$ from the TOI Catalog. We show the residuals of this comparison, $((R_\textrm{p}/R_{\star})_\textrm{TKS} - (R_\textrm{p}/R_{\star})_\textrm{TOI})/\sigma(R_\textrm{p}/R_{\star})_\textrm{TOI}$, in Figure \ref{fig:residuals}b. Here we see that 4 out of 108 TKS planets have $(R_\textrm{p}/R_{\star})_\textrm{TKS}$ measurements that are $\gtrsim$3$\sigma_\textrm{TOI}$ discrepant from $(R_\textrm{p}/R_{\star})_\textrm{TOI}$.

Upon closer inspection, we find that these discrepancies occur in instances where our careful de-trending and TTV modeling led to more reliable $R_\textrm{p}/R_{\star}$ measurements. An example of this can be seen in Figure \ref{fig:detrending}, where the initial PDC-SAP photometry of TOI-1136 displays significant variability which we remove before performing our final transit fits that include TTV modeling. Our measured $R_\textrm{p}/R_{\star}$ values are consistent within $\sim$1.5$\sigma$ of the published values reported by \citealt{Dai22} while the $R_\textrm{p}/R_{\star}$ values reported by the TOI Catalog are far more discrepant.

\subsection{Derived Planet Properties}
\label{sec:derived}

We calculate the radius of each planet in our sample ($R_\textrm{p}$) from our final weighted posterior distributions of $R_\textrm{p}/R_\star$ and our isochrone-modeled $R_\star$ values. When compared against the TOI Catalog measurements, we find that our planet radii have a lower median uncertainty, reduced from $\sigma(R_\textrm{p})/R_\textrm{p} \approx 6.7\%$ for TOI Catalog values to $3.8\%$ for TKS results. When looking at the mean uncertainty rather than the median, the difference is even more significant due to large outliers among the SPOC pipeline outputs, with mean uncertainties reduced from $\sigma(R_\textrm{p})/R_\textrm{p} \approx 16.1\%$ to $3.9\%$.

\begin{figure}[ht]
\centering
\includegraphics[width=0.45\textwidth]{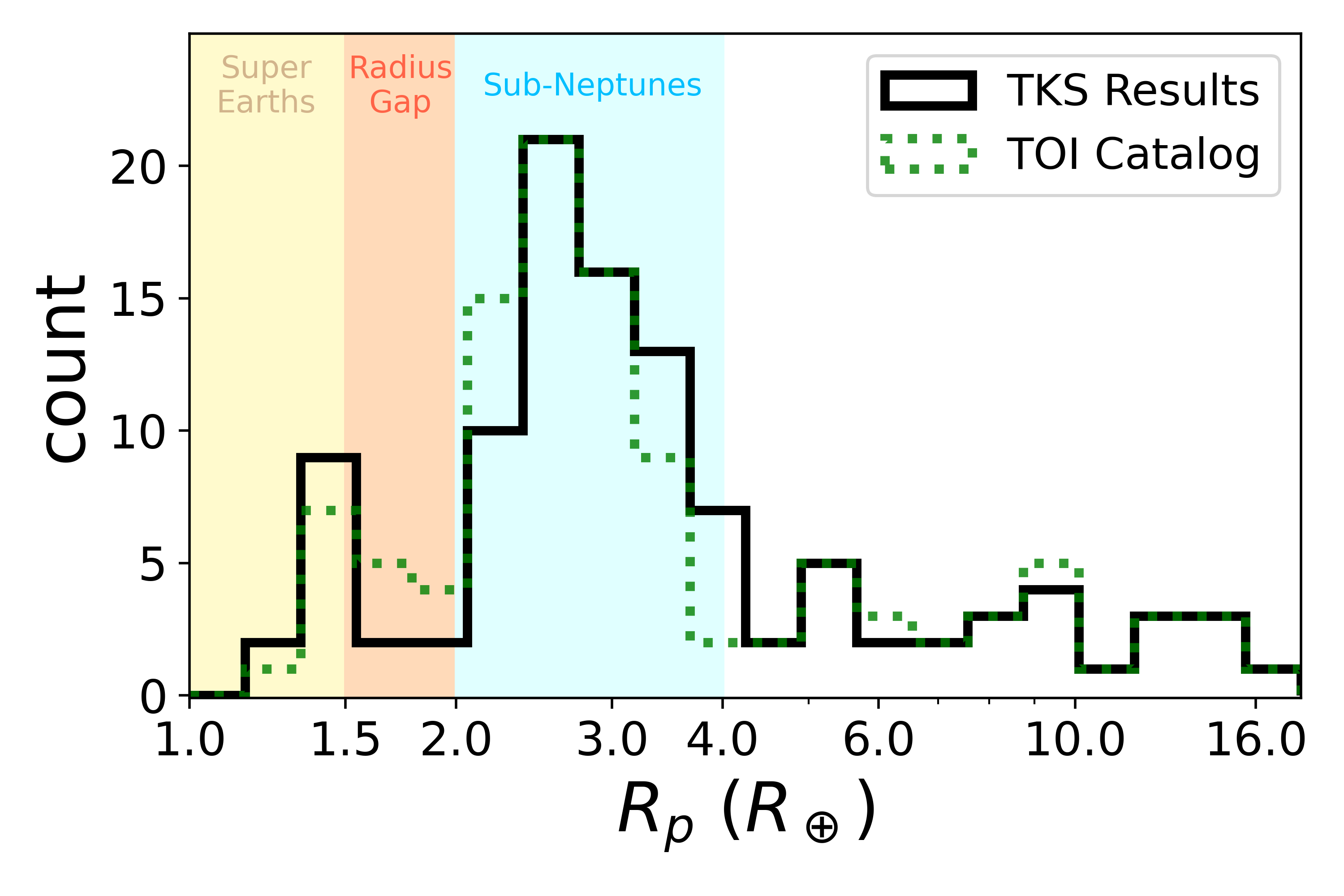}
\caption{Distribution of median planet radius measurements for the TKS sample from this work, compared against the median radii taken from the TOI Catalog (green). Shaded regions show specific planet categories, classified by planetary radius. We observe a sparser radius gap region using TKS results than is implied by radii from the TOI Catalog, likely owing to our higher precision measurements.}
\label{fig:plrad_hist}
\end{figure}

The planet radius distribution that we measure for the TKS sample is shown in Figure \ref{fig:plrad_hist}, displayed against the distribution of TKS planet radii reported by the TOI Catalog. We show the residuals of this comparison, $(R_\textrm{p, TKS} - R_\textrm{p, TOI})/\sigma(R_\textrm{p, TOI})$, in Figure \ref{fig:residuals}c. We observe similar residual distributions for both $R_\textrm{p}/R_\star$ and $R_\textrm{p}$ but far less significant scatter in the $R_{\star}$ residuals (Figure \ref{fig:residuals}a), suggesting that the uncertainties on $R_\textrm{p}$ are dominated by differences in lightcurve modeling rather than stellar characterization (\citealt{Petigura20}).

\begin{figure*}[ht]
\centering
\includegraphics[width=0.95\textwidth]{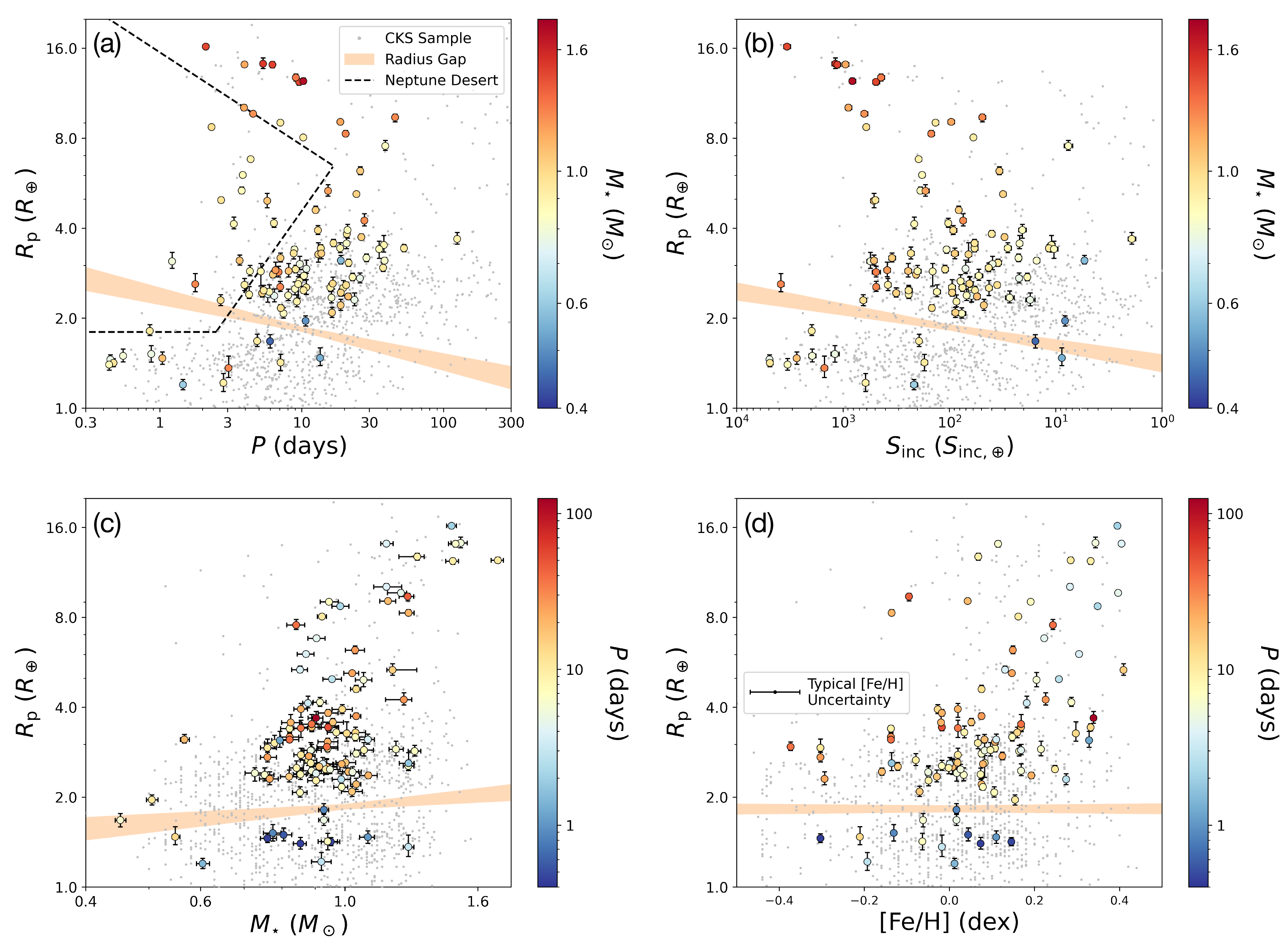}
\caption{Panel (a): radius and orbital period for all TKS planets, with stellar mass shown as a color scale. Grey background points show the distribution of planets from the CKS sample (\citealt{Petigura22}). The radius gap, as defined by \cite{Petigura22}, is shown as the orange shaded region. The Hot Neptune desert, as defined by \cite{Mazeh16}, is shown as the region marked off by black dashed lines. Panel (b): same as (a), except x-axis is $S_\textrm{inc}$. Panel (c): same as (a), except x-axis is $M_\star$ and color scale is $P$. Panel (d): same as (c), except x-axis is [Fe/H].}
\label{fig:radius-gap}
\end{figure*}

Along with planet radii, we also compute semi-major axis $a$, incident stellar flux $S_\textrm{inc}$, and equilibrium temperature $T_\textrm{eq}$ for each planet. We calculate these parameters assuming constant planet-to-star separation, which is not a valid approximation for eccentric orbits so these values should only be considered as estimates in such cases. We compute semi-major axes from Kepler's Third Law, using precise values of $P$ and our well-constrained measurements of $M_{\star}$. The median uncertainty among these measurements is $\sigma(a)/a \approx 1\%$. We also calculate $S_\textrm{inc}$ and $T_\textrm{eq}$ using the standard equations described in \cite{Johnson17}, assuming a Bond albedo of $\alpha = 0.3$ which is typical of a super-Earth-size \textit{Kepler} planet (\citealt{Demory14}). This assumption is not uniformly applicable across our sample, but it is a common simplification among similar studies such as the California-\textit{Kepler} Survey (\citealt{Johnson17}). The median uncertainties that we measure for $S_\textrm{inc}$ and $T_\textrm{eq}$ are 6.7$\%$ and 1.7$\%$, respectively, without accounting for the uncertainty in $\alpha$.

\section{Discussion}
\label{sec:discussion}

\subsection{Radius Gap}
\label{sec:radiusgap}

We note from \cite{Chontos22} that a primary goal of TKS Science Case 1A is to better understand the bimodal distribution of small planet radii first described by \cite{Fulton17}. This radius valley has been studied extensively from both \textit{Kepler} and \textit{K2} planets (see, e.g., \citealt{Berger18}; \citealt{Cloutier20}) which have improved our understanding of the physical mechanisms leading to this phenomenon (\citealt{VE18}; \citealt{Gupta19}). For the first time among TESS planets, we observe a valley-like structure between 1.5--2.0 $R_\oplus$ among the TKS sample. In Figure \ref{fig:plrad_hist}, we show the distribution of TKS planet radii and highlight the radius valley, which we find is insensitive to differences in binning. We also find this structure to be more significant among our precisely measured TKS radii from this work as compared to the TOI Catalog radii. We attribute this difference primarily to our improved constraints on $R_\textrm{p}/R_{\star}$ from our lightcurve detrending and modeling procedure (see, e.g., \citealt{Petigura20}).

The radius gap has also been found to have some reliance on various other system properties such as orbital period, incident flux (\citealt{Cloutier20}), and stellar parameters as well (\citealt{FultonPetigura18}; \citealt{Berger20b}; \citealt{Chen22}). This was recently explored by \citealt{Petigura22} for the sample of 1,246 planets in the California-\textit{Kepler} Survey (CKS; \citealt{Petigura17}), which was the \textit{Kepler}-focused predecessor to the TESS-Keck Survey. Although the TKS planet sample is only 8.4\% the size of the CKS sample, our results do suggest the existence of a radius valley in various 2D representations of radius as a function of other parameters: $R_\textrm{p}-P$, $R_\textrm{p}-S_\textrm{inc}$, $R_\textrm{p}-M_\star$, and $R_\textrm{p}$--[Fe/H] (Figure \ref{fig:radius-gap}). The orange shaded regions in these figures show the radius valley as measured by \cite{Petigura22} in the various parameter spaces, which generally agree with our results despite the CKS and TKS samples having different selection criteria. Future investigations into the radius valley may combine these samples to gain further insight to the mechanisms driving the apparent bi-modality in the distribution of small planet radii.

\subsection{Hot Neptune Desert}
\label{sec:neptunedesert}

Similar to the radius gap, the Hot Neptune Desert has been proposed as a region of $R_\textrm{p}-P$ parameter space that is sparsely populated, with few Neptune-size planets at $P \lesssim 5$ d. The mechanism that would drive such an effect is currently unknown, but this sparsity has been proposed as a natural consequence of photo-evaporation (\citealt{Lopez13}; \citealt{Owen13}). The Hot Neptune Desert region can be seen in Figure \ref{fig:radius-gap}a, outlined by boundaries defined by \cite{Mazeh16}. We find that the TKS sample includes at least 12 planets which fall firmly within this region (see, e.g., \citealt{MacDougall22}; \citealt{VanZandt22}), supplying future investigations with an increased sample size to better understand the nature of the Hot Neptune Desert. Further studies will be carried out by the TKS team to determine the compositions (TKS Science Case 1B) and atmospheric properties (TKS Science Case 3) of such TKS planets to probe this topic further with planet masses retrieved from ongoing radial velocity measurements.

\begin{figure}[ht]
\centering
\includegraphics[width=0.46\textwidth]{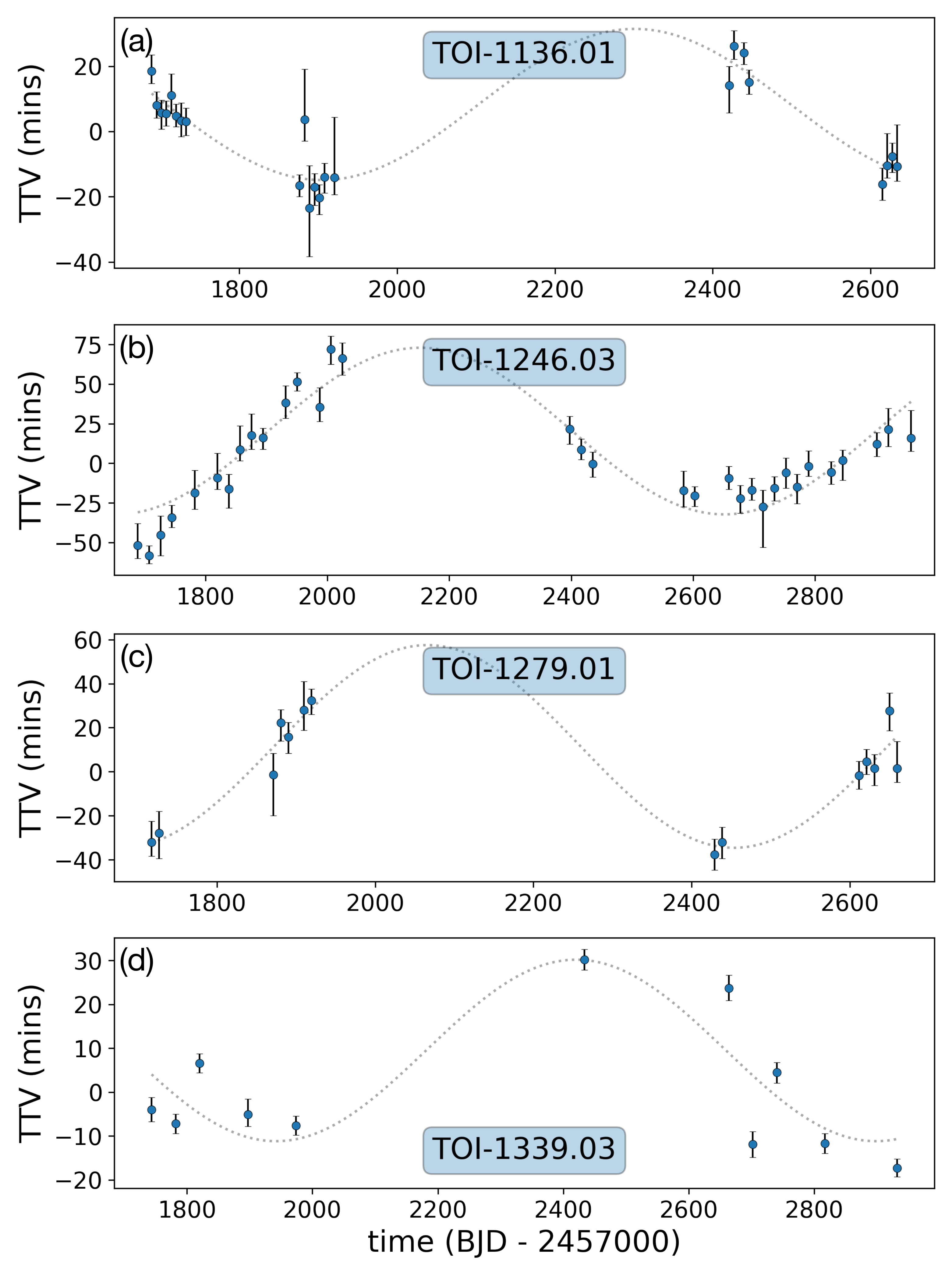}
\caption{Four examples of significant transit-timing variations detected among TKS systems, including three from multi-planet systems (panels a, b, and d) and one from a single-planet system (panel c). (Panel a) TOI-1136.01: TTV semi-amplitude $\approx$ 24 minutes, TTV signal period $\approx$ 790 days; (b) TOI-1246.03: TTV semi-amplitude $\approx$ 61 minutes, TTV signal period $\approx$ 990 days; (c) TOI-1279.01: TTV semi-amplitude $\approx$ 34 minutes, TTV signal period $\approx$ 775 days; (d) TOI-1339.03: TTV semi-amplitude $\approx$ 22 minutes, TTV signal period $\approx$ 960 days, with significant scatter likely attributed to TTV chopping. These TTVs are measured from the observed transit midpoints (fit via \texttt{exoplanet}) minus the transit midpoints calculated assuming a linear ephemeris. An estimated sinusoidal signal is shown fit to each set of TTVs, using a Lomb-Scargle periodogram and regression fit from \texttt{astropy}. These significant TTV signals likely arise from dynamical interactions between planetary companions.} 
\label{fig:ttvs}
\end{figure}

\subsection{Multis and TTVs}
\label{sec:multis}

The TKS sample was selected to contain several systems with multiple transiting planet signals, fulfilling TKS Science Case 2C. Among our 85 target systems, we report 69 with a single transiting planet signal, 11 with two transit signals, 3 with three transit signals (TOI-561, \citealt{Weiss21}; TOI-1339, \citealt{Lubin22}; TOI-2076, \citealt{Osborn22}), 1 with four transit signals (TOI-1246, \citealt{Turtelboom22}), and 1 with more than four transit signals (TOI-1136). For TOI-1136, we only report on 4 transiting planets in this work, but ongoing investigations by TKS collaborators are working towards revealing the true number of transiting planets in this system (see \citealt{Dai22}). We also measure orbital periods in some multi-planet systems that differ from those reported by the SPOC pipeline. These include TOI-561.03 (see \citealt{Lacedelli22}), TOI-266.02 (Akana Murphy et al. 2023, in prep\footnote{Period ambiguity resolved via private communications with the CHEOPS team (point of contact: Hugh Osborn)}), and TOI-1471.02 (Osborn et al. 2023, in prep). These findings help to better resolve the architectures of several multi-planet systems and are discussed in existing or upcoming literature which verify by our analysis.

We also note that several of the multi-planet systems in our sample display resonant or near-resonant orbital configurations, such as the 1:2 orbital period ratios among the inner planets in the TOI-1136 system (see \citealt{Dai22}) as well as the outer planets in the TOI-1246 system (\citealt{Turtelboom22}). Such orbital configurations can give rise to significant transit-timing variations, which we observe for TOI-1136.01 and TOI-1246.03 with TTV semi-amplitudes of $\sim$24 and $\sim$61 minutes, respectively (Figure \ref{fig:ttvs}a--b). The resonant counterparts to these two planets, TOI-1136.02 and TOI-1246.04, also display significant TTVs with similar semi-amplitudes as their companions but with possible anti-correlation and less distinguishable periodicity. We assert that our measured TTVs for a given planet are significant when $>$32$\%$ of the measured transit times differ from a linear ephemeris by $>$1$\sigma$.

The near-resonant planets in the TOI-2076 system also display possibly significant TTVs, as first noted by \cite{Osborn22}, but we do not identify a clear structure to this signal. Additionally, we observe possible TTV chopping for TOI-1339.03 (TTV semi-amplitude $\approx$ 22 minutes; Figure \ref{fig:ttvs}d) which orbits near the 3:4 mean-motion resonance with TOI-1339.02 (TTV semi-amplitude $\approx$ 10 minutes), as suggested by \cite{BadenasAgusti20}. The semi-amplitudes that we report here are only an approximation, estimated as half the difference between the maximum and minimum TTV value for a given planet.

It is also possible for transit-timing variations to arise in systems without a known transiting companion, suggesting the presence of an additional non-transiting planet. Although most single-planet orbits among the TKS sample can be described by a linear ephemeris (see \S\ref{sec:ttvs}), we identify at least one planet that displays significant TTVs without a known companion: TOI-1279.01 with a TTV semi-amplitude of $\sim$34 minutes (Figure \ref{fig:ttvs}c). At least two other single-planet systems among the TKS sample also display possible TTVs, TOI-1611.01 and TOI-1742.01, but these signals are less significant relative to the other TTV systems described above. The dynamics and multiplicities of the TKS systems discussed in this section will be examined further in future TKS studies of multi-planet systems (TKS Science Case 2C) and distant giant companions (TKS Science Case 2A).


\subsection{Eccentricity}
\label{sec:ecc}

Through importance sampling (\S\ref{sec:impsamp}), we measure posterior distributions for $e$ and $\omega$ for all planets in the TKS sample using stellar density and the results of our photometric modeling. The goal of TKS Science Case 2Bi is to characterize the eccentricities of sub-Jovian planets and better measure their underlying eccentricity distribution. Generally, however, photometric modeling alone is known to produce imprecise measurements of \{$e$, $\omega$\} on a planet-by-planet basis. Since $\omega$ is typically unconstrained from transits and $b$ is often loosely constrained for small planets due to ambiguous ingress / egress signals, strong covariances between $b$, $e$, and $\omega$ drive large uncertainties in photometrically derived eccentricities. We observe this with sub-Neptune TOI-1255.01 from the TKS sample (\citealt{MacDougall21}), for which the photometric eccentricity $e_\textrm{phot}$ was highly degenerate with $b$ until additional eccentricity constraints were introduced from radial velocity measurements to determine $e \approx 0.16$. On the other hand, we also identified the Neptune-size TKS target TOI-1272.01 (\citealt{MacDougall22}) as moderately eccentric based on $e_\textrm{phot}$ and later confirmed a self-consistent eccentricity of $e \approx 0.34$ via joint modeling with radial velocities. 

Rather than using our photometrically-derived eccentricities to draw conclusions about planets on an individual level, we are conducting an ongoing study to determine the population-level eccentricity distribution underlying the TKS sample (MacDougall et al. 2023, in prep). In this way, we are less sensitive to parameter covariances like $b$-$e$-$\omega$ and other effects like the Lucy-Sweeney bias which leads to preferentially non-circular orbit fits (\citealt{LucySweeney71}; \citealt{Eastman13}). An investigation is also underway by members of TKS to measure more precise eccentricities for planets in our sample using radial velocity measurements (Polanski et al. 2023, in prep), which will aid in breaking the degeneracy between $e$ and $b$ for lower signal-to-noise planets.

\section{Conclusions}
\label{sec:conclusions}

In this work, we characterized 85 TESS target stars presented by \cite{Chontos22} as the TESS-Keck Survey sample. Through a combination of photometry, high-resolution spectroscopy, and \textit{Gaia} parallaxes, we measured stellar properties for these targets with improved precision relative to past works. We used these stellar characterizations to facilitate in homogeneously modeling the lightcurve photometry of each TKS target, including fitting for photometric variability, estimating transit times, and measuring the transit properties of planet candidates recovered from the reduced lightcurve photometry. We characterized a total of 108 transiting planet candidates orbiting the 85 TKS target stars - the largest sample of homogeneously characterized planets from the TESS mission to date. 

In measuring the radii of these planets, we found a substantial increase in precision amongst our radii relative to the TOI Catalog ($\sigma(R_\textrm{p, TOI})/R_\textrm{p, TOI} \approx 6.7\%$ versus $\sigma(R_\textrm{p, TKS})/R_\textrm{p, TKS} \approx 3.8\%$), largely attributed to our improved accuracy in measuring $R_\textrm{p}/R_{\star}$. From our updated radii, we successfully recovered the radius gap in both one-dimensional ($R_\textrm{p}$) and two-dimensional ($R_\textrm{p}-P$, $R_\textrm{p}-S_\textrm{inc}$, $R_\textrm{p}-M_\star$, and $R_\textrm{p}$--[Fe/H]) parameter spaces. While we do not provide new models to fit the topology of the radius gap in these parameter spaces, we find that the distribution of radii in the radius gap region is broadly consistent with that of \textit{Kepler} planets. We also identify significant transit-timing variations among several TKS systems, including multi-planet systems TOI-1136, TOI-1246, and TOI-1339 as well as the single-planet system TOI-1279. These TTVs will inform future investigations into the dynamics and multiplicity of TKS systems. 

The TKS Collaboration will continue use the planetary and stellar characterizations from this work to answer key questions that remain in the field of exoplanet science, covering the topics of bulk planet compositions, system architectures and dynamics, planet atmospheres, and the effects of stellar evolution. Our photometric analysis will soon be accompanied by a radial velocity analysis of all TKS targets as well (Polanski et al. 2023, in prep), which will provide us with new planetary system characteristics such as planet masses, precise eccentricities, and discoveries of non-transiting planets. The combination of insights from the photometry described here and upcoming radial velocity measurements will allow us to more fully answer the questions laid out by the TKS Science Cases.

The stellar and planetary properties measured from this work, along with additional supporting materials, will be made available in machine-readable format on GitHub\footnote{\url{https://github.com/mason-macdougall/tks_system_properties.git}}. The TESS-Keck Survey sample adds to the legacy of in-depth surveys of host stars and their transiting exoplanets, such as the California-\textit{Kepler} Survey, which will continue to serve the community as a source of precise, homogeneously derived planetary system properties.

\begin{acknowledgments}

MM acknowledges support from the UCLA Cota-Robles Graduate Fellowship. DH acknowledges support from the Alfred P.\ Sloan Foundation, the National Aeronautics and Space Administration (80NSSC21K0652), and the National Science Foundation (AST-1717000). DD acknowledges support from the TESS Guest Investigator Program grant 80NSSC22K0185 and NASA Exoplanet Research Program grant 18-2XRP18\_2-0136. TF acknowledges support from the University of California President's Postdoctoral Fellowship Program. RAR is supported by the NSF Graduate Research Fellowship, grant No.\ DGE 1745301. PD acknowledges support from a 51 Pegasi b Postdoctoral Fellowship from the Heising-Simons Foundation. JMAM is supported by the National Science Foundation Graduate Research Fellowship Program under Grant No. DGE-1842400. JMAM acknowledges the LSSTC Data Science Fellowship Program, which is funded by LSSTC, NSF Cybertraining Grant No.\ 1829740, the Brinson Foundation, and the Moore Foundation; his participation in the program has benefited this work. 

This work was supported by a NASA Keck PI Data Award, administered by the NASA Exoplanet Science Institute. Data presented herein were obtained at the W.\ M.\ Keck Observatory from telescope time allocated to the National Aeronautics and Space Administration through the agency's scientific partnership with the California Institute of Technology and the University of California. The Observatory was made possible by the generous financial support of the W.\ M.\ Keck Foundation.

We thank the time assignment committees of the University of California, the California Institute of Technology, NASA, and the University of Hawaii for supporting the TESS-Keck Survey with observing time at Keck Observatory.  We thank NASA for funding associated with our Key Strategic Mission Support project.  We gratefully acknowledge the efforts and dedication of the Keck Observatory staff for support of HIRES and remote observing.  We recognize and acknowledge the cultural role and reverence that the summit of Maunakea has within the indigenous Hawaiian community. We are deeply grateful to have the opportunity to conduct observations from this mountain.  

This paper is based on data collected by the TESS mission. Funding for the TESS mission is provided by the NASA Explorer Program. We also acknowledge the use of public TESS data from pipelines at the TESS Science Office and at the TESS Science Processing Operations Center. This paper also includes data that are publicly available from the Mikulski Archive for Space Telescopes (MAST) at the Space Telescope Science Institute. The specific observations analyzed can be accessed via \dataset[10.17909/t9-nmc8-f686]{https://doi.org/10.17909/t9-nmc8-f686} and \dataset[10.17909/fwdt-2x66]{https://doi.org/10.17909/fwdt-2x66}. Resources supporting this work were provided by the NASA High-End Computing (HEC) Program through the NASA Advanced Supercomputing (NAS) Division at Ames Research Center for the production of the SPOC data products. 
\end{acknowledgments}

\facilities{TESS, Keck/HIRES}

\software{We made use of the following publicly available Python modules: \texttt{exoplanet} (\citealt{Foreman-Mackey2021}), \texttt{PyMC3} (\citealt{pymc16}), \texttt{theano} (\citealt{theano16}), \texttt{celerite2} (\citealt{celerite2}), \texttt{astropy} (\citealt{astropy:2013}, \citealt{astropy:2018}), \texttt{isoclassify} (\citealt{Huber17}), \texttt{lightkurve} (\citealt{lightkurve18}), \texttt{matplotlib} (\citealt{Hunter07}), \texttt{numpy} (\citealt{harris2020array}), \texttt{scipy} (\citealt{SciPy20}), \texttt{limb-darkening} (\citealt{EspinozaJordan15}), \texttt{SpecMatch-Synth} (\citealt{Petigura17b}), \texttt{SpecMatch-Emp} (\citealt{Yee17}), and \texttt{pandas} (\citealt{pandas}).}

\startlongtable 
\begin{deluxetable*}{lRRRRRRRRc}
\small
\tablecaption{Stellar Properties\label{tab:startab}}
\tabletypesize{\footnotesize}
\tablecolumns{10}
\tablewidth{0pt}
\tablehead{
	\colhead{TOI} & 
	\colhead{$m_\textrm{K}$} &
    \colhead{$\pi$} &
	\colhead{$T_\textrm{eff}$} &
	\colhead{[Fe/H]} &
    \colhead{$R_\star$} & 
	\colhead{$M_\star$} & 
	\colhead{$\rho_\star$} & 
    \colhead{age} &
    \colhead{SpecMatch} \\
    \colhead{} & 
    \colhead{mag} & 
    \colhead{mas} &
	\colhead{K} &
	\colhead{dex} & 
    \colhead{$R_\odot$} & 
	\colhead{$M_\odot$} & 
	\colhead{$g$ $cm^{-3}$} & 
    \colhead{Gyr} &
    \colhead{}
}
\startdata
260 &	6.55 &	49.51 &	4050\substack{+35 \\ -30} &	-0.21\substack{+0.06 \\ -0.09} &	0.55\substack{+0.01 \\ -0.01} &	0.55\substack{+0.01 \\ -0.01} &	4.7\substack{+0.14 \\ -0.2} &	15.9\substack{+2.8 \\ -5.5} &	emp \\
266 &	8.45 &	9.8 &	5617\substack{+74 \\ -73} &	0.01\substack{+0.06 \\ -0.06} &	0.96\substack{+0.02 \\ -0.02} &	0.94\substack{+0.03 \\ -0.03} &	1.48\substack{+0.11 \\ -0.1} &	6.0\substack{+2.4 \\ -2.3} &	syn \\
329 &	9.68 &	3.49 &	5613\substack{+78 \\ -73} &	0.2\substack{+0.06 \\ -0.06} &	1.53\substack{+0.03 \\ -0.03} &	1.07\substack{+0.03 \\ -0.03} &	0.41\substack{+0.03 \\ -0.03} &	8.7\substack{+1.0 \\ -1.0} &	syn \\
465 &	9.34 &	8.16 &	4986\substack{+57 \\ -56} &	0.3\substack{+0.06 \\ -0.06} &	0.83\substack{+0.01 \\ -0.01} &	0.87\substack{+0.02 \\ -0.03} &	2.15\substack{+0.15 \\ -0.14} &	5.6\substack{+3.9 \\ -3.2} &	syn \\
469 &	7.59 &	14.64 &	5322\substack{+72 \\ -70} &	0.33\substack{+0.06 \\ -0.06} &	0.98\substack{+0.02 \\ -0.02} &	0.95\substack{+0.03 \\ -0.03} &	1.39\substack{+0.1 \\ -0.1} &	8.6\substack{+2.8 \\ -2.7} &	syn \\
480 &	6.01 &	18.31 &	6173\substack{+89 \\ -118} &	0.16\substack{+0.08 \\ -0.07} &	1.49\substack{+0.04 \\ -0.03} &	1.28\substack{+0.03 \\ -0.03} &	0.54\substack{+0.04 \\ -0.05} &	2.7\substack{+0.6 \\ -0.5} &	syn \\
509 &	6.88 &	20.39 &	5512\substack{+71 \\ -72} &	0.11\substack{+0.05 \\ -0.06} &	0.96\substack{+0.02 \\ -0.02} &	0.94\substack{+0.03 \\ -0.03} &	1.49\substack{+0.11 \\ -0.1} &	6.6\substack{+2.5 \\ -2.4} &	syn \\
554 &	5.71 &	21.89 &	6352\substack{+84 \\ -76} &	-0.02\substack{+0.05 \\ -0.06} &	1.42\substack{+0.02 \\ -0.02} &	1.25\substack{+0.02 \\ -0.02} &	0.61\substack{+0.04 \\ -0.03} &	2.3\substack{+0.4 \\ -0.4} &	syn \\
561 &	8.39 &	11.63 &	5342\substack{+57 \\ -54} &	-0.3\substack{+0.06 \\ -0.06} &	0.86\substack{+0.01 \\ -0.01} &	0.76\substack{+0.02 \\ -0.02} &	1.69\substack{+0.11 \\ -0.09} &	16.1\substack{+2.4 \\ -3.1} &	syn \\
669 &	9.13 &	6.97 &	5597\substack{+71 \\ -72} &	0.0\substack{+0.06 \\ -0.06} &	0.99\substack{+0.02 \\ -0.02} &	0.92\substack{+0.03 \\ -0.03} &	1.34\substack{+0.1 \\ -0.09} &	7.9\substack{+2.3 \\ -2.2} &	syn \\
1136 &	8.03 &	11.8 &	5805\substack{+57 \\ -57} &	0.08\substack{+0.05 \\ -0.06} &	0.96\substack{+0.01 \\ -0.01} &	1.04\substack{+0.02 \\ -0.02} &	1.66\substack{+0.06 \\ -0.07} &	0.7\substack{+0.9 \\ -0.4} &	syn \\
1173 &	9.13 &	7.53 &	5414\substack{+62 \\ -65} &	0.19\substack{+0.06 \\ -0.06} &	0.94\substack{+0.02 \\ -0.01} &	0.95\substack{+0.03 \\ -0.03} &	1.61\substack{+0.11 \\ -0.1} &	5.9\substack{+2.4 \\ -2.4} &	syn \\
1174 &	8.96 &	10.54 &	5157\substack{+59 \\ -57} &	-0.0\substack{+0.06 \\ -0.06} &	0.78\substack{+0.03 \\ -0.02} &	0.84\substack{+0.02 \\ -0.02} &	2.44\substack{+0.18 \\ -0.25} &	2.7\substack{+3.6 \\ -1.9} &	syn \\
1180 &	8.59 &	13.86 &	4803\substack{+53 \\ -52} &	0.07\substack{+0.08 \\ -0.09} &	0.74\substack{+0.02 \\ -0.02} &	0.78\substack{+0.02 \\ -0.03} &	2.66\substack{+0.22 \\ -0.28} &	5.0\substack{+6.3 \\ -3.6} &	emp \\
1181 &	9.22 &	3.27 &	6045\substack{+88 \\ -84} &	0.39\substack{+0.05 \\ -0.06} &	1.93\substack{+0.04 \\ -0.04} &	1.46\substack{+0.02 \\ -0.03} &	0.28\substack{+0.02 \\ -0.02} &	2.4\substack{+0.3 \\ -0.3} &	syn \\
1184 &	8.32 &	17.04 &	4616\substack{+47 \\ -48} &	0.02\substack{+0.08 \\ -0.09} &	0.7\substack{+0.02 \\ -0.02} &	0.73\substack{+0.02 \\ -0.03} &	3.0\substack{+0.24 \\ -0.27} &	5.3\substack{+6.6 \\ -3.9} &	emp \\
1194 &	9.34 &	6.65 &	5393\substack{+64 \\ -62} &	0.35\substack{+0.06 \\ -0.06} &	0.96\substack{+0.02 \\ -0.01} &	0.98\substack{+0.03 \\ -0.03} &	1.55\substack{+0.1 \\ -0.1} &	4.9\substack{+2.3 \\ -2.1} &	syn \\
1244 &	9.42 &	9.72 &	4721\substack{+55 \\ -54} &	0.03\substack{+0.09 \\ -0.09} &	0.72\substack{+0.02 \\ -0.02} &	0.75\substack{+0.02 \\ -0.03} &	2.78\substack{+0.24 \\ -0.27} &	5.9\substack{+6.8 \\ -4.2} &	emp \\
1246 &	9.91 &	5.87 &	5213\substack{+70 \\ -67} &	0.17\substack{+0.06 \\ -0.06} &	0.86\substack{+0.02 \\ -0.02} &	0.89\substack{+0.03 \\ -0.03} &	1.92\substack{+0.16 \\ -0.15} &	6.3\substack{+3.6 \\ -3.2} &	syn \\
1247 &	7.5 &	13.51 &	5697\substack{+75 \\ -72} &	-0.12\substack{+0.06 \\ -0.06} &	1.07\substack{+0.02 \\ -0.02} &	0.91\substack{+0.03 \\ -0.03} &	1.03\substack{+0.07 \\ -0.07} &	9.8\substack{+2.0 \\ -2.0} &	syn \\
1248 &	9.87 &	5.9 &	5205\substack{+56 \\ -61} &	0.22\substack{+0.06 \\ -0.06} &	0.87\substack{+0.01 \\ -0.01} &	0.9\substack{+0.03 \\ -0.03} &	1.93\substack{+0.13 \\ -0.12} &	5.5\substack{+3.0 \\ -2.7} &	syn \\
1249 &	9.13 &	7.14 &	5496\substack{+68 \\ -66} &	0.3\substack{+0.06 \\ -0.06} &	0.98\substack{+0.02 \\ -0.02} &	1.01\substack{+0.03 \\ -0.03} &	1.5\substack{+0.11 \\ -0.1} &	4.0\substack{+2.3 \\ -2.0} &	syn \\
1255 &	7.92 &	15.13 &	5216\substack{+52 \\ -52} &	0.25\substack{+0.08 \\ -0.05} &	0.84\substack{+0.01 \\ -0.01} &	0.93\substack{+0.02 \\ -0.02} &	2.19\substack{+0.08 \\ -0.09} &	1.2\substack{+1.5 \\ -0.8} &	syn \\
1269 &	9.89 &	5.78 &	5499\substack{+63 \\ -60} &	-0.05\substack{+0.06 \\ -0.06} &	0.85\substack{+0.01 \\ -0.01} &	0.9\substack{+0.02 \\ -0.03} &	2.04\substack{+0.12 \\ -0.12} &	3.1\substack{+2.3 \\ -1.8} &	syn \\
1272 &	9.7 &	7.24 &	5065\substack{+52 \\ -50} &	0.18\substack{+0.05 \\ -0.06} &	0.79\substack{+0.01 \\ -0.01} &	0.88\substack{+0.01 \\ -0.02} &	2.46\substack{+0.08 \\ -0.09} &	1.1\substack{+1.6 \\ -0.8} &	syn \\
1279 &	8.89 &	9.31 &	5457\substack{+66 \\ -67} &	-0.08\substack{+0.06 \\ -0.06} &	0.84\substack{+0.01 \\ -0.01} &	0.88\substack{+0.03 \\ -0.03} &	2.05\substack{+0.13 \\ -0.13} &	4.2\substack{+2.5 \\ -2.2} &	syn \\
1288 &	8.78 &	8.68 &	5388\substack{+68 \\ -66} &	0.26\substack{+0.06 \\ -0.06} &	0.96\substack{+0.02 \\ -0.02} &	0.96\substack{+0.03 \\ -0.03} &	1.52\substack{+0.11 \\ -0.1} &	6.5\substack{+2.6 \\ -2.5} &	syn \\
1294 &	9.96 &	2.98 &	5718\substack{+58 \\ -77} &	0.28\substack{+0.06 \\ -0.06} &	1.56\substack{+0.03 \\ -0.03} &	1.16\substack{+0.06 \\ -0.05} &	0.43\substack{+0.03 \\ -0.03} &	6.2\substack{+1.4 \\ -1.4} &	syn \\
1296 &	9.74 &	3.06 &	5567\substack{+77 \\ -72} &	0.4\substack{+0.06 \\ -0.02} &	1.7\substack{+0.03 \\ -0.03} &	1.16\substack{+0.03 \\ -0.02} &	0.33\substack{+0.02 \\ -0.02} &	7.1\substack{+0.6 \\ -0.7} &	syn \\
1298 &	10.01 &	3.11 &	5752\substack{+68 \\ -98} &	0.4\substack{+0.06 \\ -0.06} &	1.45\substack{+0.03 \\ -0.02} &	1.22\substack{+0.02 \\ -0.06} &	0.55\substack{+0.03 \\ -0.04} &	4.3\substack{+1.7 \\ -0.6} &	syn \\
1339 &	7.18 &	18.62 &	5385\substack{+69 \\ -64} &	-0.14\substack{+0.06 \\ -0.06} &	0.93\substack{+0.02 \\ -0.02} &	0.82\substack{+0.03 \\ -0.03} &	1.41\substack{+0.1 \\ -0.09} &	13.9\substack{+2.8 \\ -2.7} &	syn \\
1347 &	9.62 &	6.75 &	5493\substack{+53 \\ -55} &	0.02\substack{+0.06 \\ -0.05} &	0.84\substack{+0.01 \\ -0.01} &	0.93\substack{+0.02 \\ -0.02} &	2.19\substack{+0.08 \\ -0.08} &	0.8\substack{+1.1 \\ -0.6} &	syn \\
1386 &	9.09 &	6.78 &	5734\substack{+75 \\ -76} &	0.15\substack{+0.06 \\ -0.06} &	1.02\substack{+0.02 \\ -0.02} &	1.04\substack{+0.03 \\ -0.04} &	1.35\substack{+0.09 \\ -0.09} &	3.2\substack{+1.9 \\ -1.6} &	syn \\
1410 &	8.58 &	13.72 &	4635\substack{+52 \\ -51} &	0.33\substack{+0.08 \\ -0.09} &	0.76\substack{+0.02 \\ -0.02} &	0.79\substack{+0.02 \\ -0.02} &	2.57\substack{+0.22 \\ -0.28} &	6.6\substack{+6.6 \\ -4.6} &	emp \\
1411 &	7.25 &	30.76 &	4115\substack{+41 \\ -45} &	0.01\substack{+0.07 \\ -0.07} &	0.61\substack{+0.02 \\ -0.02} &	0.6\substack{+0.01 \\ -0.01} &	3.63\substack{+0.23 \\ -0.23} &	9.6\substack{+6.6 \\ -6.3} &	emp \\
1422 &	9.19 &	6.42 &	5810\substack{+81 \\ -80} &	-0.03\substack{+0.06 \\ -0.06} &	1.02\substack{+0.02 \\ -0.02} &	0.99\substack{+0.04 \\ -0.04} &	1.29\substack{+0.09 \\ -0.09} &	4.6\substack{+2.0 \\ -1.8} &	syn \\
1430 &	7.08 &	24.26 &	5086\substack{+62 \\ -59} &	0.11\substack{+0.06 \\ -0.06} &	0.8\substack{+0.03 \\ -0.02} &	0.85\substack{+0.02 \\ -0.02} &	2.34\substack{+0.19 \\ -0.26} &	3.4\substack{+4.3 \\ -2.4} &	syn \\
1436 &	9.72 &	7.57 &	5029\substack{+58 \\ -57} &	-0.13\substack{+0.06 \\ -0.06} &	0.75\substack{+0.03 \\ -0.02} &	0.77\substack{+0.02 \\ -0.03} &	2.58\substack{+0.28 \\ -0.32} &	5.9\substack{+5.9 \\ -4.0} &	syn \\
1437 &	7.82 &	9.65 &	5985\substack{+82 \\ -81} &	-0.16\substack{+0.03 \\ -0.05} &	1.26\substack{+0.02 \\ -0.02} &	1.01\substack{+0.04 \\ -0.04} &	0.71\substack{+0.05 \\ -0.05} &	6.7\substack{+1.4 \\ -1.3} &	syn \\
1438 &	9.09 &	9.02 &	5237\substack{+55 \\ -55} &	0.08\substack{+0.05 \\ -0.05} &	0.82\substack{+0.01 \\ -0.01} &	0.89\substack{+0.02 \\ -0.02} &	2.29\substack{+0.1 \\ -0.13} &	2.2\substack{+2.2 \\ -1.5} &	syn \\
1439 &	9.05 &	4.31 &	5840\substack{+63 \\ -64} &	0.23\substack{+0.06 \\ -0.07} &	1.62\substack{+0.03 \\ -0.03} &	1.23\substack{+0.04 \\ -0.08} &	0.4\substack{+0.02 \\ -0.03} &	4.6\substack{+1.6 \\ -0.7} &	syn \\
1443 &	8.67 &	11.63 &	5223\substack{+58 \\ -57} &	-0.29\substack{+0.06 \\ -0.06} &	0.75\substack{+0.03 \\ -0.03} &	0.77\substack{+0.02 \\ -0.03} &	2.49\substack{+0.33 \\ -0.35} &	7.4\substack{+5.9 \\ -4.8} &	syn \\
1444 &	9.06 &	7.94 &	5460\substack{+66 \\ -66} &	0.15\substack{+0.06 \\ -0.06} &	0.91\substack{+0.02 \\ -0.01} &	0.95\substack{+0.03 \\ -0.03} &	1.76\substack{+0.12 \\ -0.11} &	3.8\substack{+2.5 \\ -2.1} &	syn \\
1451 &	8.07 &	10.85 &	5800\substack{+77 \\ -78} &	0.02\substack{+0.06 \\ -0.06} &	1.02\substack{+0.02 \\ -0.02} &	1.0\substack{+0.04 \\ -0.03} &	1.34\substack{+0.09 \\ -0.09} &	4.0\substack{+1.9 \\ -1.7} &	syn \\
1456 &	7.24 &	12.37 &	6127\substack{+87 \\ -79} &	0.04\substack{+0.06 \\ -0.06} &	1.27\substack{+0.02 \\ -0.02} &	1.16\substack{+0.03 \\ -0.03} &	0.8\substack{+0.05 \\ -0.04} &	2.9\substack{+0.9 \\ -0.8} &	syn \\
1467 &	8.57 &	26.68 &	3775\substack{+22 \\ -20} &	-0.06\substack{+0.04 \\ -0.05} &	0.46\substack{+0.01 \\ -0.01} &	0.45\substack{+0.01 \\ -0.01} &	6.65\substack{+0.22 \\ -0.24} &	13.4\substack{+4.4 \\ -6.4} &	emp \\
1471 &	7.56 &	14.78 &	5648\substack{+74 \\ -70} &	-0.02\substack{+0.06 \\ -0.06} &	0.96\substack{+0.02 \\ -0.02} &	0.94\substack{+0.03 \\ -0.03} &	1.49\substack{+0.11 \\ -0.1} &	5.6\substack{+2.3 \\ -2.2} &	syn \\
1472 &	9.28 &	8.18 &	5132\substack{+54 \\ -53} &	0.29\substack{+0.06 \\ -0.06} &	0.84\substack{+0.01 \\ -0.01} &	0.92\substack{+0.02 \\ -0.02} &	2.18\substack{+0.1 \\ -0.13} &	2.3\substack{+2.5 \\ -1.5} &	syn \\
1473 &	7.39 &	14.76 &	5934\substack{+74 \\ -78} &	-0.05\substack{+0.06 \\ -0.05} &	1.01\substack{+0.02 \\ -0.02} &	1.03\substack{+0.03 \\ -0.04} &	1.41\substack{+0.09 \\ -0.09} &	2.4\substack{+1.6 \\ -1.3} &	syn \\
1601 &	9.19 &	2.94 &	5982\substack{+78 \\ -64} &	0.34\substack{+0.05 \\ -0.06} &	2.2\substack{+0.06 \\ -0.06} &	1.51\substack{+0.03 \\ -0.05} &	0.2\substack{+0.02 \\ -0.02} &	2.4\substack{+0.4 \\ -0.2} &	syn \\
1611 &	6.31 &	35.31 &	5106\substack{+66 \\ -63} &	-0.03\substack{+0.06 \\ -0.06} &	0.78\substack{+0.03 \\ -0.03} &	0.82\substack{+0.02 \\ -0.03} &	2.38\substack{+0.25 \\ -0.31} &	4.7\substack{+5.1 \\ -3.2} &	syn \\
1669 &	8.46 &	8.96 &	5497\substack{+73 \\ -72} &	0.27\substack{+0.06 \\ -0.06} &	1.06\substack{+0.02 \\ -0.02} &	0.99\substack{+0.03 \\ -0.03} &	1.17\substack{+0.08 \\ -0.08} &	7.7\substack{+2.2 \\ -2.1} &	syn \\
1691 &	8.54 &	8.94 &	5641\substack{+80 \\ -74} &	0.05\substack{+0.06 \\ -0.06} &	1.01\substack{+0.02 \\ -0.02} &	0.96\substack{+0.04 \\ -0.03} &	1.3\substack{+0.1 \\ -0.09} &	6.6\substack{+2.3 \\ -2.2} &	syn \\
1694 &	9.43 &	7.99 &	5058\substack{+60 \\ -55} &	0.13\substack{+0.05 \\ -0.06} &	0.8\substack{+0.01 \\ -0.01} &	0.85\substack{+0.02 \\ -0.02} &	2.3\substack{+0.14 \\ -0.15} &	4.3\substack{+3.5 \\ -2.6} &	syn \\
1710 &	7.96 &	12.28 &	5684\substack{+63 \\ -67} &	0.15\substack{+0.04 \\ -0.05} &	0.96\substack{+0.02 \\ -0.02} &	1.03\substack{+0.02 \\ -0.03} &	1.64\substack{+0.08 \\ -0.09} &	1.5\substack{+1.6 \\ -1.0} &	syn \\
1716 &	7.93 &	9.57 &	5877\substack{+78 \\ -78} &	0.1\substack{+0.06 \\ -0.06} &	1.22\substack{+0.02 \\ -0.02} &	1.07\substack{+0.04 \\ -0.04} &	0.83\substack{+0.06 \\ -0.06} &	5.4\substack{+1.4 \\ -1.3} &	syn \\
1723 &	8.13 &	9.9 &	5742\substack{+82 \\ -80} &	0.16\substack{+0.05 \\ -0.06} &	1.09\substack{+0.02 \\ -0.02} &	1.04\substack{+0.04 \\ -0.04} &	1.13\substack{+0.09 \\ -0.08} &	4.8\substack{+1.9 \\ -1.8} &	syn \\
1726 &	5.26 &	44.61 &	5634\substack{+62 \\ -64} &	0.08\substack{+0.05 \\ -0.05} &	0.92\substack{+0.01 \\ -0.01} &	0.99\substack{+0.02 \\ -0.03} &	1.8\substack{+0.08 \\ -0.09} &	1.5\substack{+1.6 \\ -1.0} &	syn \\
1736 &	7.28 &	11.21 &	5636\substack{+86 \\ -79} &	0.15\substack{+0.06 \\ -0.06} &	1.43\substack{+0.03 \\ -0.02} &	1.04\substack{+0.04 \\ -0.03} &	0.5\substack{+0.04 \\ -0.03} &	9.2\substack{+1.3 \\ -1.4} &	syn \\
1742 &	7.34 &	13.69 &	5814\substack{+75 \\ -72} &	0.19\substack{+0.06 \\ -0.06} &	1.13\substack{+0.02 \\ -0.02} &	1.09\substack{+0.03 \\ -0.03} &	1.05\substack{+0.07 \\ -0.06} &	3.9\substack{+1.5 \\ -1.3} &	syn \\
1751 &	7.93 &	8.79 &	5970\substack{+85 \\ -83} &	-0.37\substack{+0.06 \\ -0.06} &	1.31\substack{+0.03 \\ -0.02} &	0.94\substack{+0.04 \\ -0.03} &	0.58\substack{+0.04 \\ -0.04} &	9.0\substack{+1.4 \\ -1.4} &	syn \\
1753 &	10.18 &	4.38 &	5620\substack{+75 \\ -71} &	0.03\substack{+0.06 \\ -0.06} &	0.97\substack{+0.02 \\ -0.02} &	0.95\substack{+0.03 \\ -0.03} &	1.44\substack{+0.12 \\ -0.11} &	5.9\substack{+2.4 \\ -2.3} &	syn \\
1758 &	8.81 &	10.32 &	5150\substack{+64 \\ -59} &	-0.02\substack{+0.06 \\ -0.06} &	0.82\substack{+0.01 \\ -0.01} &	0.83\substack{+0.03 \\ -0.03} &	2.13\substack{+0.15 \\ -0.15} &	7.5\substack{+3.7 \\ -3.2} &	syn \\
1759 &	7.93 &	24.93 &	3930\substack{+32 \\ -29} &	0.08\substack{+0.04 \\ -0.04} &	0.57\substack{+0.01 \\ -0.01} &	0.57\substack{+0.01 \\ -0.01} &	4.18\substack{+0.15 \\ -0.17} &	15.1\substack{+3.4 \\ -5.4} &	emp \\
1775 &	9.72 &	6.67 &	5283\substack{+50 \\ -51} &	0.16\substack{+0.03 \\ -0.05} &	0.83\substack{+0.01 \\ -0.01} &	0.92\substack{+0.01 \\ -0.02} &	2.27\substack{+0.07 \\ -0.08} &	0.8\substack{+1.1 \\ -0.5} &	syn \\
1776 &	6.69 &	22.37 &	5784\substack{+74 \\ -71} &	-0.19\substack{+0.06 \\ -0.06} &	0.94\substack{+0.02 \\ -0.02} &	0.92\substack{+0.03 \\ -0.03} &	1.57\substack{+0.11 \\ -0.1} &	5.1\substack{+2.2 \\ -2.0} &	syn \\
1778 &	7.62 &	10.03 &	6007\substack{+83 \\ -87} &	0.22\substack{+0.06 \\ -0.06} &	1.32\substack{+0.02 \\ -0.01} &	1.2\substack{+0.03 \\ -0.03} &	0.73\substack{+0.03 \\ -0.05} &	3.1\substack{+0.9 \\ -0.7} &	syn \\
1794 &	8.69 &	6.39 &	5631\substack{+77 \\ -73} &	0.03\substack{+0.05 \\ -0.06} &	1.32\substack{+0.02 \\ -0.02} &	0.97\substack{+0.03 \\ -0.03} &	0.59\substack{+0.04 \\ -0.04} &	10.8\substack{+1.5 \\ -1.5} &	syn \\
1797 &	7.78 &	12.12 &	5907\substack{+65 \\ -71} &	0.11\substack{+0.06 \\ -0.06} &	1.03\substack{+0.02 \\ -0.02} &	1.08\substack{+0.03 \\ -0.03} &	1.38\substack{+0.07 \\ -0.08} &	1.3\substack{+1.2 \\ -0.8} &	syn \\
1798 &	9.24 &	8.81 &	5106\substack{+53 \\ -55} &	0.07\substack{+0.06 \\ -0.05} &	0.79\substack{+0.01 \\ -0.01} &	0.85\substack{+0.02 \\ -0.02} &	2.42\substack{+0.1 \\ -0.13} &	2.6\substack{+2.6 \\ -1.7} &	syn \\
1799 &	7.38 &	16.07 &	5697\substack{+74 \\ -68} &	-0.06\substack{+0.06 \\ -0.05} &	0.95\substack{+0.02 \\ -0.02} &	0.94\substack{+0.03 \\ -0.03} &	1.52\substack{+0.1 \\ -0.1} &	5.1\substack{+2.2 \\ -2.0} &	syn \\
1801 &	7.8 &	32.57 &	3747\substack{+24 \\ -23} &	0.15\substack{+0.05 \\ -0.04} &	0.51\substack{+0.01 \\ -0.01} &	0.51\substack{+0.01 \\ -0.01} &	5.23\substack{+0.23 \\ -0.23} &	13.3\substack{+4.6 \\ -6.6} &	emp \\
1807 &	7.57 &	23.46 &	4914\substack{+60 \\ -57} &	0.04\substack{+0.09 \\ -0.1} &	0.75\substack{+0.02 \\ -0.02} &	0.8\substack{+0.02 \\ -0.03} &	2.72\substack{+0.18 \\ -0.2} &	2.1\substack{+3.4 \\ -1.5} &	emp \\
1823 &	8.33 &	13.93 &	4926\substack{+57 \\ -56} &	0.24\substack{+0.09 \\ -0.09} &	0.8\substack{+0.03 \\ -0.03} &	0.84\substack{+0.03 \\ -0.03} &	2.24\substack{+0.25 \\ -0.31} &	6.3\substack{+6.3 \\ -4.4} &	emp \\
1824 &	7.76 &	16.8 &	5165\substack{+56 \\ -56} &	0.12\substack{+0.06 \\ -0.06} &	0.81\substack{+0.01 \\ -0.01} &	0.88\substack{+0.02 \\ -0.02} &	2.3\substack{+0.1 \\ -0.13} &	2.5\substack{+2.5 \\ -1.6} &	syn \\
1836 &	8.53 &	5.19 &	6237\substack{+91 \\ -68} &	-0.14\substack{+0.07 \\ -0.03} &	1.65\substack{+0.03 \\ -0.03} &	1.25\substack{+0.02 \\ -0.04} &	0.39\substack{+0.02 \\ -0.02} &	3.2\substack{+0.7 \\ -0.3} &	syn \\
1842 &	8.45 &	4.45 &	6039\substack{+93 \\ -88} &	0.33\substack{+0.06 \\ -0.06} &	2.03\substack{+0.04 \\ -0.04} &	1.46\substack{+0.03 \\ -0.04} &	0.24\substack{+0.02 \\ -0.02} &	2.5\substack{+0.4 \\ -0.3} &	syn \\
1898 &	6.66 &	12.52 &	6241\substack{+88 \\ -84} &	-0.1\substack{+0.06 \\ -0.07} &	1.61\substack{+0.03 \\ -0.03} &	1.25\substack{+0.03 \\ -0.04} &	0.41\substack{+0.03 \\ -0.02} &	3.1\substack{+0.6 \\ -0.3} &	syn \\
2019 &	8.6 &	5.02 &	5625\substack{+77 \\ -77} &	0.41\substack{+0.06 \\ -0.06} &	1.75\substack{+0.03 \\ -0.03} &	1.18\substack{+0.11 \\ -0.02} &	0.31\substack{+0.03 \\ -0.02} &	6.4\substack{+0.6 \\ -1.9} &	syn \\
2045 &	9.85 &	2.7 &	6045\substack{+75 \\ -68} &	0.07\substack{+0.06 \\ -0.04} &	1.78\substack{+0.04 \\ -0.04} &	1.29\substack{+0.03 \\ -0.08} &	0.32\substack{+0.02 \\ -0.02} &	3.4\substack{+1.2 \\ -0.4} &	syn \\
2076 &	7.12 &	23.83 &	5191\substack{+61 \\ -57} &	0.02\substack{+0.06 \\ -0.06} &	0.8\substack{+0.03 \\ -0.02} &	0.86\substack{+0.02 \\ -0.02} &	2.36\substack{+0.19 \\ -0.25} &	2.7\substack{+3.6 \\ -1.9} &	syn \\
2088 &	9.52 &	7.88 &	5080\substack{+55 \\ -54} &	0.34\substack{+0.06 \\ -0.06} &	0.85\substack{+0.03 \\ -0.02} &	0.9\substack{+0.02 \\ -0.03} &	2.05\substack{+0.2 \\ -0.26} &	4.5\substack{+4.6 \\ -3.1} &	syn \\
2114 &	9.09 &	3.12 &	6359\substack{+88 \\ -85} &	0.11\substack{+0.05 \\ -0.06} &	2.1\substack{+0.04 \\ -0.04} &	1.48\substack{+0.03 \\ -0.04} &	0.23\substack{+0.01 \\ -0.01} &	2.0\substack{+0.2 \\ -0.2} &	syn \\
2128 &	5.82 &	27.31 &	5967\substack{+85 \\ -84} &	-0.07\substack{+0.05 \\ -0.06} &	1.12\substack{+0.02 \\ -0.02} &	1.03\substack{+0.04 \\ -0.04} &	1.02\substack{+0.08 \\ -0.08} &	4.7\substack{+1.7 \\ -1.5} &	syn \\
2145 &	7.76 &	4.42 &	6200\substack{+83 \\ -81} &	0.29\substack{+0.06 \\ -0.05} &	2.75\substack{+0.06 \\ -0.05} &	1.72\substack{+0.03 \\ -0.04} &	0.11\substack{+0.01 \\ -0.01} &	1.6\substack{+0.2 \\ -0.1} &	syn \\
\enddata
\tablecomments{Properties of 85 planet hosting stars from the TKS sample (\citealt{Chontos22}). $m_\textrm{K}$ is 2MASS $K$-band apparent magnitude and $\pi$ is \textit{Gaia} DR2 parallax. $T_\textrm{eff}$, [Fe/H], $R_\star$, $M_\star$, $\rho_\star$ and age were derived using \texttt{isoclassify} isochrone modeling (\citealt{Huber17}) as described in \S\ref{sec:isochrone}. The given values reflect median measurements with upper and lower uncertainties. The last column denotes which \texttt{SpecMatch} method was used to derive the initial stellar property inputs to \texttt{isoclassify}: "syn" -- \texttt{SpecMatch-Synthetic} or "emp" -- \texttt{SpecMatch-Empirical} (see \S\ref{sec:specmatch}). }
\end{deluxetable*}

\startlongtable 
\begin{deluxetable*}{lRRRRRRRc}
\small
\tablecaption{Planet Properties\label{tab:planettab}}
\tabletypesize{\footnotesize}
\tablecolumns{8}
\tablewidth{0pt}
\tablehead{
	\colhead{Planet} & 
	\colhead{$P$} &
    \colhead{$t_{0}$} &
	\colhead{$R_\textrm{p}/R_\star$} &
    \colhead{$R_\textrm{p}$} & 
	\colhead{$T_{14}$} &
	\colhead{$T_\textrm{circ}$} &
    \colhead{$a$} &
    \colhead{TTVs?} \\
    \colhead{} & 
    \colhead{d} & 
    \colhead{BTJD} &
	\colhead{$\%$} & 
    \colhead{$R_\oplus$} &
	\colhead{hr} &
	\colhead{hr} &
	\colhead{au} &
	\colhead{y/n}
}
\startdata
260.01 &	13.475832\substack{+4.5e-05 \\ -5.6e-05} &	1392.2944\substack{+0.0021 \\ -0.0023} &	2.47\substack{+0.2 \\ -0.14} &	1.47\substack{+0.12 \\ -0.08} &	2.0\substack{+0.14 \\ -0.13} &	2.96\substack{+0.03 \\ -0.04} &	0.0907\substack{+0.0005 \\ -0.0003} &	n \\
266.01 &	10.751013\substack{+7.2e-05 \\ -7e-05} &	1393.0862\substack{+0.0035 \\ -0.0032} &	2.42\substack{+0.14 \\ -0.12} &	2.54\substack{+0.16 \\ -0.14} &	3.0\substack{+0.16 \\ -0.13} &	4.04\substack{+0.1 \\ -0.09} &	0.0933\substack{+0.0012 \\ -0.0011} &	n \\
266.02 &	19.605464\substack{+0.000228 \\ -0.000237} &	1398.2927\substack{+0.0087 \\ -0.0059} &	2.41\substack{+0.14 \\ -0.13} &	2.52\substack{+0.15 \\ -0.14} &	4.38\substack{+0.27 \\ -0.21} &	4.94\substack{+0.12 \\ -0.11} &	0.1393\substack{+0.0017 \\ -0.0016} &	n \\
329.01 &	5.70442\substack{+0.000116 \\ -0.000101} &	2090.7934\substack{+0.0043 \\ -0.0057} &	2.96\substack{+0.16 \\ -0.14} &	4.94\substack{+0.29 \\ -0.26} &	4.21\substack{+0.21 \\ -0.18} &	5.04\substack{+0.13 \\ -0.11} &	0.0639\substack{+0.0006 \\ -0.0006} &	n \\
465.01 &	3.836162\substack{+2e-06 \\ -2e-06} &	1414.1361\substack{+0.0004 \\ -0.0003} &	6.68\substack{+0.09 \\ -0.07} &	6.03\substack{+0.13 \\ -0.12} &	2.4\substack{+0.02 \\ -0.02} &	2.64\substack{+0.06 \\ -0.06} &	0.0458\substack{+0.0004 \\ -0.0005} &	n \\
469.01 &	13.630829\substack{+2.6e-05 \\ -2.6e-05} &	1474.5691\substack{+0.0011 \\ -0.0011} &	3.19\substack{+0.07 \\ -0.06} &	3.43\substack{+0.1 \\ -0.09} &	4.33\substack{+0.05 \\ -0.04} &	4.5\substack{+0.11 \\ -0.11} &	0.1096\substack{+0.0013 \\ -0.0012} &	n \\
480.01 &	6.865906\substack{+1.7e-05 \\ -1.9e-05} &	1469.5655\substack{+0.0016 \\ -0.0015} &	1.76\substack{+0.06 \\ -0.05} &	2.85\substack{+0.13 \\ -0.1} &	3.57\substack{+0.06 \\ -0.06} &	4.84\substack{+0.13 \\ -0.15} &	0.0767\substack{+0.0006 \\ -0.0006} &	n \\
509.01 &	9.058805\substack{+1.8e-05 \\ -1.6e-05} &	1494.4467\substack{+0.001 \\ -0.001} &	2.79\substack{+0.07 \\ -0.06} &	2.92\substack{+0.09 \\ -0.08} &	3.7\substack{+0.04 \\ -0.04} &	3.82\substack{+0.09 \\ -0.09} &	0.0833\substack{+0.001 \\ -0.0009} &	n \\
509.02 &	21.402464\substack{+0.001794 \\ -0.001919} &	1504.1438\substack{+0.067 \\ -0.0622} &	2.93\substack{+0.07 \\ -0.06} &	3.06\substack{+0.09 \\ -0.08} &	5.42\substack{+0.08 \\ -0.06} &	5.1\substack{+0.12 \\ -0.12} &	0.1478\substack{+0.0018 \\ -0.0017} &	n \\
554.01 &	3.04405\substack{+9e-06 \\ -8e-06} &	1438.4733\substack{+0.002 \\ -0.0025} &	0.88\substack{+0.08 \\ -0.06} &	1.36\substack{+0.13 \\ -0.1} &	1.35\substack{+0.06 \\ -0.06} &	3.52\substack{+0.08 \\ -0.05} &	0.0443\substack{+0.0003 \\ -0.0003} &	n \\
554.02 &	7.049157\substack{+1.3e-05 \\ -1.4e-05} &	1442.6179\substack{+0.0016 \\ -0.0016} &	1.65\substack{+0.07 \\ -0.05} &	2.55\substack{+0.11 \\ -0.08} &	3.35\substack{+0.06 \\ -0.04} &	4.68\substack{+0.1 \\ -0.07} &	0.0775\substack{+0.0005 \\ -0.0005} &	n \\
561.01 &	0.4465691\substack{+4e-07 \\ -4e-07} &	1517.9451\substack{+0.0007 \\ -0.0008} &	1.56\substack{+0.05 \\ -0.04} &	1.46\substack{+0.05 \\ -0.05} &	1.35\substack{+0.02 \\ -0.02} &	1.36\substack{+0.03 \\ -0.03} &	0.0104\substack{+0.0001 \\ -0.0001} &	n \\
561.02 &	10.778858\substack{+3.3e-05 \\ -3.5e-05} &	1527.0608\substack{+0.0025 \\ -0.0023} &	3.13\substack{+0.21 \\ -0.09} &	2.93\substack{+0.2 \\ -0.1} &	3.83\substack{+0.16 \\ -0.11} &	3.89\substack{+0.08 \\ -0.07} &	0.0872\substack{+0.0009 \\ -0.0007} &	n \\
561.03 &	25.712443\substack{+0.000111 \\ -0.000122} &	1521.8835\substack{+0.0038 \\ -0.0036} &	2.91\substack{+0.11 \\ -0.09} &	2.72\substack{+0.11 \\ -0.1} &	5.19\substack{+0.17 \\ -0.12} &	5.19\substack{+0.11 \\ -0.09} &	0.1556\substack{+0.0017 \\ -0.0013} &	n \\
669.01 &	3.945154\substack{+1.3e-05 \\ -1.1e-05} &	1546.1417\substack{+0.0021 \\ -0.003} &	2.4\substack{+0.11 \\ -0.09} &	2.59\substack{+0.13 \\ -0.11} &	2.87\substack{+0.1 \\ -0.09} &	2.99\substack{+0.07 \\ -0.07} &	0.0476\substack{+0.0006 \\ -0.0005} &	n \\
1136.01 &	6.258767\substack{+1.5e-05 \\ -1.4e-05} &	1688.7067\substack{+0.0013 \\ -0.0012} &	2.65\substack{+0.06 \\ -0.05} &	2.77\substack{+0.08 \\ -0.07} &	3.49\substack{+0.05 \\ -0.04} &	3.25\substack{+0.04 \\ -0.04} &	0.0673\substack{+0.0004 \\ -0.0004} &	y \\
1136.02 &	12.518595\substack{+1.5e-05 \\ -1.5e-05} &	1686.0634\substack{+0.0005 \\ -0.0006} &	4.43\substack{+0.13 \\ -0.08} &	4.62\substack{+0.15 \\ -0.1} &	4.13\substack{+0.07 \\ -0.04} &	4.17\substack{+0.05 \\ -0.06} &	0.1069\substack{+0.0006 \\ -0.0007} &	y \\
1136.03 &	18.797505\substack{+0.000141 \\ -0.000121} &	1697.799\substack{+0.004 \\ -0.0045} &	2.44\substack{+0.09 \\ -0.08} &	2.55\substack{+0.1 \\ -0.09} &	4.46\substack{+0.11 \\ -0.1} &	4.68\substack{+0.06 \\ -0.06} &	0.1402\substack{+0.0008 \\ -0.0009} &	y \\
1136.04 &	26.317767\substack{+4.7e-05 \\ -4.7e-05} &	1699.3798\substack{+0.0011 \\ -0.0011} &	3.58\substack{+0.06 \\ -0.06} &	3.74\substack{+0.09 \\ -0.08} &	5.18\substack{+0.05 \\ -0.05} &	5.3\substack{+0.07 \\ -0.07} &	0.1755\substack{+0.001 \\ -0.0011} &	y \\
1173.01 &	7.064397\substack{+3e-06 \\ -3e-06} &	1688.7154\substack{+0.0002 \\ -0.0002} &	8.83\substack{+0.06 \\ -0.06} &	9.02\substack{+0.16 \\ -0.15} &	2.56\substack{+0.02 \\ -0.02} &	3.63\substack{+0.08 \\ -0.08} &	0.0707\substack{+0.0008 \\ -0.0007} &	n \\
1174.01 &	8.9534\substack{+4.1e-05 \\ -3.9e-05} &	1690.0626\substack{+0.0022 \\ -0.0026} &	2.93\substack{+0.13 \\ -0.1} &	2.51\substack{+0.14 \\ -0.11} &	3.0\substack{+0.12 \\ -0.1} &	3.23\substack{+0.08 \\ -0.11} &	0.0797\substack{+0.0006 \\ -0.0007} &	n \\
1180.01 &	9.686755\substack{+1.2e-05 \\ -1.2e-05} &	1691.0488\substack{+0.0009 \\ -0.0009} &	3.75\substack{+0.2 \\ -0.12} &	3.04\substack{+0.19 \\ -0.13} &	2.76\substack{+0.08 \\ -0.05} &	3.25\substack{+0.09 \\ -0.11} &	0.0818\substack{+0.0008 \\ -0.0009} &	n \\
1181.01 &	2.1031937\substack{+3e-07 \\ -3e-07} &	1957.8213\substack{+0.0001 \\ -0.0001} &	7.68\substack{+0.02 \\ -0.02} &	16.2\substack{+0.33 \\ -0.34} &	4.12\substack{+0.01 \\ -0.01} &	4.33\substack{+0.09 \\ -0.08} &	0.0364\substack{+0.0002 \\ -0.0002} &	n \\
1184.01 &	5.748431\substack{+3e-06 \\ -3e-06} &	1684.3594\substack{+0.0005 \\ -0.0005} &	3.17\substack{+0.09 \\ -0.16} &	2.41\substack{+0.1 \\ -0.14} &	1.88\substack{+0.03 \\ -0.03} &	2.61\substack{+0.07 \\ -0.08} &	0.0565\substack{+0.0006 \\ -0.0006} &	n \\
1194.01 &	2.310645\substack{+1e-06 \\ -1e-06} &	1684.9235\substack{+0.0002 \\ -0.0002} &	8.32\substack{+0.06 \\ -0.06} &	8.72\substack{+0.16 \\ -0.15} &	1.63\substack{+0.01 \\ -0.01} &	2.53\substack{+0.06 \\ -0.05} &	0.034\substack{+0.0003 \\ -0.0003} &	n \\
1244.01 &	6.400317\substack{+8e-06 \\ -8e-06} &	1684.9474\substack{+0.0008 \\ -0.0008} &	3.01\substack{+0.14 \\ -0.1} &	2.38\substack{+0.13 \\ -0.11} &	2.22\substack{+0.04 \\ -0.04} &	2.77\substack{+0.08 \\ -0.09} &	0.0614\substack{+0.0007 \\ -0.0007} &	n \\
1246.01 &	4.30744\substack{+4e-06 \\ -3e-06} &	1686.5661\substack{+0.0007 \\ -0.0008} &	3.05\substack{+0.12 \\ -0.08} &	2.87\substack{+0.13 \\ -0.1} &	2.18\substack{+0.04 \\ -0.03} &	2.75\substack{+0.08 \\ -0.07} &	0.0498\substack{+0.0006 \\ -0.0006} &	n \\
1246.02 &	5.904141\substack{+1.3e-05 \\ -1.4e-05} &	1683.4664\substack{+0.0016 \\ -0.0016} &	2.6\substack{+0.14 \\ -0.08} &	2.45\substack{+0.14 \\ -0.09} &	2.89\substack{+0.07 \\ -0.07} &	3.04\substack{+0.08 \\ -0.08} &	0.0615\substack{+0.0007 \\ -0.0007} &	n \\
1246.03 &	18.654958\substack{+4.9e-05 \\ -4.7e-05} &	1688.9746\substack{+0.002 \\ -0.002} &	3.64\substack{+0.14 \\ -0.1} &	3.43\substack{+0.15 \\ -0.12} &	3.95\substack{+0.1 \\ -0.08} &	4.5\substack{+0.12 \\ -0.12} &	0.1324\substack{+0.0015 \\ -0.0015} &	y \\
1246.04 &	37.925389\substack{+0.000149 \\ -0.000156} &	1700.6956\substack{+0.0033 \\ -0.0034} &	3.73\substack{+0.27 \\ -0.18} &	3.51\substack{+0.27 \\ -0.18} &	3.54\substack{+0.16 \\ -0.12} &	5.71\substack{+0.16 \\ -0.15} &	0.2124\substack{+0.0024 \\ -0.0024} &	y \\
1247.01 &	15.923365\substack{+3.4e-05 \\ -3.6e-05} &	1687.6494\substack{+0.0014 \\ -0.0014} &	2.16\substack{+0.06 \\ -0.05} &	2.53\substack{+0.08 \\ -0.07} &	4.33\substack{+0.07 \\ -0.07} &	5.18\substack{+0.12 \\ -0.11} &	0.1202\substack{+0.0014 \\ -0.0013} &	n \\
1248.01 &	4.360156\substack{+1e-06 \\ -1e-06} &	1687.1212\substack{+0.0002 \\ -0.0002} &	7.19\substack{+0.05 \\ -0.06} &	6.81\substack{+0.12 \\ -0.12} &	2.31\substack{+0.01 \\ -0.01} &	2.87\substack{+0.06 \\ -0.06} &	0.0505\substack{+0.0005 \\ -0.0005} &	n \\
1249.01 &	13.079163\substack{+6.3e-05 \\ -6.3e-05} &	1694.3786\substack{+0.0027 \\ -0.0029} &	3.06\substack{+0.28 \\ -0.17} &	3.27\substack{+0.31 \\ -0.19} &	3.15\substack{+0.26 \\ -0.15} &	4.32\substack{+0.1 \\ -0.1} &	0.1089\substack{+0.0011 \\ -0.0011} &	n \\
1255.01 &	10.288889\substack{+4e-06 \\ -4e-06} &	1691.6544\substack{+0.0004 \\ -0.0003} &	2.7\substack{+0.07 \\ -0.05} &	2.48\substack{+0.07 \\ -0.05} &	1.53\substack{+0.02 \\ -0.01} &	3.5\substack{+0.04 \\ -0.05} &	0.0905\substack{+0.0005 \\ -0.0006} &	n \\
1269.01 &	4.252989\substack{+6e-06 \\ -6e-06} &	1686.6058\substack{+0.0009 \\ -0.0013} &	2.58\substack{+0.16 \\ -0.07} &	2.4\substack{+0.15 \\ -0.07} &	2.62\substack{+0.07 \\ -0.05} &	2.67\substack{+0.05 \\ -0.05} &	0.0496\substack{+0.0005 \\ -0.0005} &	n \\
1269.02 &	9.237875\substack{+2e-05 \\ -2e-05} &	1685.9772\substack{+0.0022 \\ -0.0017} &	2.45\substack{+0.12 \\ -0.09} &	2.27\substack{+0.12 \\ -0.09} &	2.72\substack{+0.07 \\ -0.07} &	3.45\substack{+0.07 \\ -0.07} &	0.0833\substack{+0.0008 \\ -0.0008} &	n \\
1272.01 &	3.315979\substack{+6e-06 \\ -5e-06} &	1713.0315\substack{+0.0006 \\ -0.0006} &	4.77\substack{+0.25 \\ -0.17} &	4.13\substack{+0.23 \\ -0.15} &	1.57\substack{+0.06 \\ -0.05} &	2.36\substack{+0.03 \\ -0.03} &	0.0417\substack{+0.0002 \\ -0.0003} &	n \\
1279.01 &	9.614216\substack{+3.7e-05 \\ -3.5e-05} &	1717.4777\substack{+0.0026 \\ -0.0027} &	2.9\substack{+0.14 \\ -0.11} &	2.66\substack{+0.14 \\ -0.11} &	2.76\substack{+0.11 \\ -0.09} &	3.5\substack{+0.07 \\ -0.07} &	0.0847\substack{+0.0009 \\ -0.0008} &	y \\
1288.01 &	2.699831\substack{+1e-06 \\ -1e-06} &	1712.3587\substack{+0.0002 \\ -0.0002} &	4.75\substack{+0.06 \\ -0.05} &	4.97\substack{+0.11 \\ -0.1} &	2.38\substack{+0.01 \\ -0.01} &	2.59\substack{+0.06 \\ -0.06} &	0.0374\substack{+0.0004 \\ -0.0004} &	n \\
1294.01 &	3.915289\substack{+1.6e-05 \\ -1.6e-05} &	2393.0074\substack{+0.0007 \\ -0.0007} &	5.96\substack{+0.08 \\ -0.08} &	10.12\substack{+0.24 \\ -0.22} &	3.03\substack{+0.05 \\ -0.05} &	4.51\substack{+0.1 \\ -0.12} &	0.051\substack{+0.0009 \\ -0.0007} &	n \\
1296.01 &	3.944373\substack{+1e-06 \\ -1e-06} &	1930.7553\substack{+0.0002 \\ -0.0002} &	7.6\substack{+0.04 \\ -0.04} &	14.12\substack{+0.28 \\ -0.26} &	4.86\substack{+0.01 \\ -0.01} &	5.04\substack{+0.12 \\ -0.1} &	0.0513\substack{+0.0004 \\ -0.0003} &	n \\
1298.01 &	4.537142\substack{+2e-06 \\ -2e-06} &	1934.1225\substack{+0.0003 \\ -0.0003} &	6.09\substack{+0.05 \\ -0.04} &	9.66\substack{+0.2 \\ -0.17} &	4.03\substack{+0.02 \\ -0.02} &	4.38\substack{+0.09 \\ -0.1} &	0.0573\substack{+0.0004 \\ -0.0009} &	n \\
1339.01 &	8.880322\substack{+3e-06 \\ -3e-06} &	1715.3558\substack{+0.0003 \\ -0.0002} &	3.33\substack{+0.05 \\ -0.07} &	3.4\substack{+0.08 \\ -0.09} &	3.07\substack{+0.02 \\ -0.02} &	3.89\substack{+0.09 \\ -0.08} &	0.0786\substack{+0.0009 \\ -0.0008} &	n \\
1339.02 &	28.579984\substack{+2.4e-05 \\ -2.4e-05} &	1726.0547\substack{+0.0006 \\ -0.0006} &	3.12\substack{+0.08 \\ -0.07} &	3.18\substack{+0.1 \\ -0.09} &	4.47\substack{+0.05 \\ -0.04} &	5.73\substack{+0.14 \\ -0.12} &	0.1714\substack{+0.0019 \\ -0.0018} &	y \\
1339.03 &	38.352218\substack{+4.2e-05 \\ -4.1e-05} &	1743.556\substack{+0.0008 \\ -0.0007} &	3.06\substack{+0.05 \\ -0.04} &	3.12\substack{+0.07 \\ -0.07} &	5.51\substack{+0.05 \\ -0.04} &	6.31\substack{+0.15 \\ -0.13} &	0.2085\substack{+0.0023 \\ -0.0022} &	y \\
1347.01 &	0.8474247\substack{+4e-07 \\ -4e-07} &	1683.5587\substack{+0.0003 \\ -0.0003} &	1.98\substack{+0.09 \\ -0.06} &	1.81\substack{+0.09 \\ -0.06} &	0.88\substack{+0.02 \\ -0.01} &	1.53\substack{+0.02 \\ -0.02} &	0.0171\substack{+0.0001 \\ -0.0001} &	n \\
1347.02 &	4.84195\substack{+1.3e-05 \\ -1.2e-05} &	1688.1903\substack{+0.0017 \\ -0.002} &	1.83\substack{+0.09 \\ -0.07} &	1.68\substack{+0.09 \\ -0.07} &	2.24\substack{+0.09 \\ -0.08} &	2.7\substack{+0.03 \\ -0.03} &	0.0546\substack{+0.0003 \\ -0.0003} &	n \\
1386.01 &	25.840653\substack{+0.003119 \\ -0.003192} &	1752.3213\substack{+0.0022 \\ -0.0022} &	5.57\substack{+0.16 \\ -0.12} &	6.22\substack{+0.21 \\ -0.17} &	5.67\substack{+0.15 \\ -0.12} &	5.75\substack{+0.13 \\ -0.13} &	0.1731\substack{+0.0019 \\ -0.002} &	n \\
1410.01 &	1.216876\substack{+1e-06 \\ -1e-06} &	1739.7293\substack{+0.0006 \\ -0.0006} &	3.76\substack{+0.24 \\ -0.17} &	3.1\substack{+0.22 \\ -0.16} &	1.11\substack{+0.05 \\ -0.04} &	1.66\substack{+0.05 \\ -0.06} &	0.0207\substack{+0.0002 \\ -0.0002} &	n \\
1411.01 &	1.452053\substack{+2e-06 \\ -2e-06} &	1739.474\substack{+0.0005 \\ -0.0005} &	1.79\substack{+0.05 \\ -0.04} &	1.2\substack{+0.05 \\ -0.04} &	1.51\substack{+0.02 \\ -0.02} &	1.53\substack{+0.03 \\ -0.03} &	0.0212\substack{+0.0002 \\ -0.0002} &	n \\
1422.01 &	13.001782\substack{+5e-05 \\ -4.2e-05} &	1745.9141\substack{+0.0018 \\ -0.0016} &	3.53\substack{+0.12 \\ -0.09} &	3.94\substack{+0.15 \\ -0.13} &	4.44\substack{+0.11 \\ -0.09} &	4.55\substack{+0.11 \\ -0.11} &	0.1079\substack{+0.0013 \\ -0.0013} &	n \\
1430.01 &	7.434098\substack{+9e-06 \\ -9e-06} &	1690.779\substack{+0.0009 \\ -0.0009} &	2.38\substack{+0.07 \\ -0.05} &	2.08\substack{+0.09 \\ -0.07} &	2.67\substack{+0.04 \\ -0.04} &	3.06\substack{+0.08 \\ -0.11} &	0.0707\substack{+0.0006 \\ -0.0006} &	n \\
1436.01 &	0.867617\substack{+2e-06 \\ -3e-06} &	1711.8961\substack{+0.0012 \\ -0.0011} &	1.87\substack{+0.1 \\ -0.09} &	1.52\substack{+0.1 \\ -0.09} &	1.32\substack{+0.05 \\ -0.05} &	1.45\substack{+0.05 \\ -0.06} &	0.0164\substack{+0.0002 \\ -0.0002} &	n \\
1437.01 &	18.840952\substack{+7.7e-05 \\ -7.7e-05} &	1700.7335\substack{+0.0029 \\ -0.003} &	1.77\substack{+0.04 \\ -0.03} &	2.43\substack{+0.07 \\ -0.06} &	6.19\substack{+0.12 \\ -0.11} &	6.19\substack{+0.15 \\ -0.14} &	0.1391\substack{+0.0017 \\ -0.0017} &	n \\
1438.01 &	5.139659\substack{+5e-06 \\ -5e-06} &	1683.6266\substack{+0.0007 \\ -0.0007} &	3.22\substack{+0.2 \\ -0.34} &	2.87\substack{+0.18 \\ -0.3} &	1.01\substack{+0.06 \\ -0.1} &	2.75\substack{+0.04 \\ -0.05} &	0.056\substack{+0.0004 \\ -0.0004} &	n \\
1438.02 &	9.428074\substack{+1.6e-05 \\ -1.5e-05} &	1689.9152\substack{+0.0014 \\ -0.0014} &	2.92\substack{+0.17 \\ -0.3} &	2.6\substack{+0.15 \\ -0.27} &	1.56\substack{+0.11 \\ -0.1} &	3.36\substack{+0.05 \\ -0.06} &	0.0839\substack{+0.0006 \\ -0.0007} &	n \\
1439.01 &	27.643869\substack{+0.000133 \\ -0.000138} &	1703.4782\substack{+0.004 \\ -0.0039} &	2.4\substack{+0.12 \\ -0.09} &	4.24\substack{+0.23 \\ -0.17} &	5.63\substack{+0.21 \\ -0.14} &	8.52\substack{+0.17 \\ -0.2} &	0.1918\substack{+0.0018 \\ -0.004} &	n \\
1443.01 &	23.540714\substack{+6.6e-05 \\ -6.2e-05} &	1693.2443\substack{+0.0023 \\ -0.0025} &	2.81\substack{+0.09 \\ -0.07} &	2.3\substack{+0.13 \\ -0.1} &	4.26\substack{+0.08 \\ -0.07} &	4.43\substack{+0.2 \\ -0.21} &	0.1471\substack{+0.0015 \\ -0.0017} &	n \\
1444.01 &	0.4702737\substack{+3e-07 \\ -3e-07} &	1711.3666\substack{+0.0005 \\ -0.0008} &	1.43\substack{+0.04 \\ -0.03} &	1.42\substack{+0.05 \\ -0.04} &	1.28\substack{+0.02 \\ -0.02} &	1.36\substack{+0.03 \\ -0.03} &	0.0116\substack{+0.0001 \\ -0.0001} &	n \\
1451.01 &	16.537903\substack{+6.9e-05 \\ -6.3e-05} &	1694.3138\substack{+0.0021 \\ -0.0029} &	2.36\substack{+0.11 \\ -0.08} &	2.61\substack{+0.13 \\ -0.1} &	3.09\substack{+0.09 \\ -0.07} &	4.82\substack{+0.11 \\ -0.11} &	0.1271\substack{+0.0015 \\ -0.0014} &	n \\
1456.01 &	18.711796\substack{+1.4e-05 \\ -1.4e-05} &	1692.2609\substack{+0.0005 \\ -0.0005} &	6.57\substack{+0.06 \\ -0.05} &	9.07\substack{+0.19 \\ -0.18} &	6.17\substack{+0.03 \\ -0.03} &	6.21\substack{+0.14 \\ -0.12} &	0.1451\substack{+0.0013 \\ -0.0014} &	y \\
1467.01 &	5.971148\substack{+1.1e-05 \\ -1e-05} &	1766.9883\substack{+0.0013 \\ -0.0022} &	3.37\substack{+0.17 \\ -0.16} &	1.68\substack{+0.09 \\ -0.08} &	1.58\substack{+0.07 \\ -0.05} &	2.03\substack{+0.02 \\ -0.02} &	0.0494\substack{+0.0003 \\ -0.0003} &	n \\
1471.01 &	20.772891\substack{+4.8e-05 \\ -5e-05} &	1767.422\substack{+0.0014 \\ -0.0014} &	3.66\substack{+0.07 \\ -0.06} &	3.83\substack{+0.1 \\ -0.09} &	4.94\substack{+0.06 \\ -0.05} &	5.08\substack{+0.12 \\ -0.11} &	0.145\substack{+0.0017 \\ -0.0017} &	n \\
1471.02 &	52.563553\substack{+0.000166 \\ -0.000185} &	1779.1909\substack{+0.0016 \\ -0.0014} &	3.28\substack{+0.07 \\ -0.06} &	3.43\substack{+0.1 \\ -0.09} &	6.57\substack{+0.08 \\ -0.06} &	6.9\substack{+0.17 \\ -0.15} &	0.2692\substack{+0.0032 \\ -0.0031} &	n \\
1472.01 &	6.363386\substack{+9e-06 \\ -8e-06} &	1765.6096\substack{+0.0012 \\ -0.0013} &	4.55\substack{+0.15 \\ -0.12} &	4.16\substack{+0.16 \\ -0.13} &	2.61\substack{+0.06 \\ -0.05} &	3.04\substack{+0.05 \\ -0.06} &	0.0653\substack{+0.0005 \\ -0.0005} &	n \\
1473.01 &	5.254479\substack{+1.2e-05 \\ -1.3e-05} &	1769.7834\substack{+0.0026 \\ -0.0019} &	2.21\substack{+0.08 \\ -0.06} &	2.43\substack{+0.1 \\ -0.07} &	3.24\substack{+0.1 \\ -0.07} &	3.23\substack{+0.07 \\ -0.07} &	0.0597\substack{+0.0006 \\ -0.0007} &	n \\
1601.01 &	5.333298\substack{+0.000889 \\ -0.000968} &	1793.2752\substack{+0.0017 \\ -0.0016} &	5.93\substack{+0.21 \\ -0.13} &	14.2\substack{+0.65 \\ -0.51} &	6.49\substack{+0.13 \\ -0.11} &	6.5\substack{+0.18 \\ -0.18} &	0.0685\substack{+0.0005 \\ -0.0008} &	n \\
1611.01 &	16.201708\substack{+1.3e-05 \\ -1.2e-05} &	1796.495\substack{+0.0006 \\ -0.0006} &	2.74\substack{+0.09 \\ -0.05} &	2.34\substack{+0.12 \\ -0.09} &	2.9\substack{+0.03 \\ -0.03} &	3.96\substack{+0.14 \\ -0.17} &	0.1173\substack{+0.0011 \\ -0.0012} &	y \\
1669.01 &	2.680055\substack{+3e-06 \\ -3e-06} &	1816.9447\substack{+0.0009 \\ -0.0008} &	1.99\substack{+0.08 \\ -0.07} &	2.3\substack{+0.1 \\ -0.09} &	1.93\substack{+0.04 \\ -0.03} &	2.75\substack{+0.07 \\ -0.06} &	0.0376\substack{+0.0004 \\ -0.0004} &	n \\
1691.01 &	16.7369\substack{+3.2e-05 \\ -3e-05} &	1818.0907\substack{+0.0013 \\ -0.0014} &	3.24\substack{+0.08 \\ -0.05} &	3.57\substack{+0.11 \\ -0.09} &	4.97\substack{+0.07 \\ -0.05} &	4.92\substack{+0.12 \\ -0.12} &	0.1262\substack{+0.0016 \\ -0.0015} &	n \\
1694.01 &	3.770137\substack{+8.8e-05 \\ -8.9e-05} &	1817.2664\substack{+0.0006 \\ -0.0006} &	6.09\substack{+0.14 \\ -0.1} &	5.34\substack{+0.15 \\ -0.12} &	2.87\substack{+0.04 \\ -0.03} &	2.55\substack{+0.05 \\ -0.05} &	0.045\substack{+0.0004 \\ -0.0004} &	n \\
1710.01 &	24.283384\substack{+1.9e-05 \\ -2e-05} &	1836.9629\substack{+0.0005 \\ -0.0005} &	4.98\substack{+0.04 \\ -0.04} &	5.2\substack{+0.1 \\ -0.09} &	5.26\substack{+0.03 \\ -0.02} &	5.25\substack{+0.09 \\ -0.1} &	0.1655\substack{+0.0013 \\ -0.0016} &	n \\
1716.01 &	8.082366\substack{+3.5e-05 \\ -3.6e-05} &	1843.8553\substack{+0.0034 \\ -0.0031} &	2.17\substack{+0.12 \\ -0.09} &	2.88\substack{+0.17 \\ -0.13} &	3.51\substack{+0.17 \\ -0.16} &	4.45\substack{+0.1 \\ -0.1} &	0.0806\substack{+0.0009 \\ -0.0009} &	n \\
1723.01 &	13.726468\substack{+0.000386 \\ -0.000398} &	1852.7027\substack{+0.0027 \\ -0.0023} &	2.78\substack{+0.11 \\ -0.09} &	3.29\substack{+0.14 \\ -0.13} &	4.4\substack{+0.22 \\ -0.22} &	4.81\substack{+0.12 \\ -0.12} &	0.1136\substack{+0.0013 \\ -0.0013} &	n \\
1726.01 &	7.107941\substack{+7e-06 \\ -6e-06} &	1845.3735\substack{+0.0005 \\ -0.0006} &	2.16\substack{+0.05 \\ -0.04} &	2.16\substack{+0.06 \\ -0.05} &	3.24\substack{+0.04 \\ -0.03} &	3.29\substack{+0.05 \\ -0.06} &	0.0721\substack{+0.0005 \\ -0.0007} &	n \\
1726.02 &	20.543827\substack{+2.1e-05 \\ -2.1e-05} &	1844.0589\substack{+0.0006 \\ -0.0006} &	2.58\substack{+0.04 \\ -0.04} &	2.58\substack{+0.06 \\ -0.06} &	4.07\substack{+0.04 \\ -0.04} &	4.7\substack{+0.07 \\ -0.08} &	0.1462\substack{+0.001 \\ -0.0013} &	n \\
1736.01 &	7.073092\substack{+1.7e-05 \\ -1.7e-05} &	1792.7939\substack{+0.0018 \\ -0.0019} &	2.04\substack{+0.07 \\ -0.05} &	3.18\substack{+0.12 \\ -0.1} &	3.98\substack{+0.08 \\ -0.08} &	5.04\substack{+0.12 \\ -0.11} &	0.073\substack{+0.0009 \\ -0.0008} &	n \\
1742.01 &	21.269084\substack{+5.4e-05 \\ -5.1e-05} &	1725.352\substack{+0.0018 \\ -0.0022} &	1.92\substack{+0.03 \\ -0.03} &	2.37\substack{+0.06 \\ -0.05} &	6.29\substack{+0.07 \\ -0.07} &	5.66\substack{+0.12 \\ -0.11} &	0.1544\substack{+0.0016 \\ -0.0016} &	y \\
1751.01 &	37.468555\substack{+0.000141 \\ -0.000138} &	1733.6315\substack{+0.0028 \\ -0.0029} &	2.06\substack{+0.07 \\ -0.04} &	2.95\substack{+0.11 \\ -0.08} &	7.65\substack{+0.16 \\ -0.1} &	8.35\substack{+0.21 \\ -0.19} &	0.2145\substack{+0.0028 \\ -0.0025} &	n \\
1753.01 &	5.384623\substack{+9e-06 \\ -1.1e-05} &	1684.504\substack{+0.0016 \\ -0.0013} &	2.34\substack{+0.08 \\ -0.06} &	2.48\substack{+0.1 \\ -0.08} &	3.18\substack{+0.05 \\ -0.05} &	3.23\substack{+0.09 \\ -0.08} &	0.0591\substack{+0.0007 \\ -0.0007} &	n \\
1758.01 &	20.705061\substack{+4.8e-05 \\ -4.8e-05} &	1806.6981\substack{+0.0015 \\ -0.0016} &	4.0\substack{+0.25 \\ -0.15} &	3.56\substack{+0.23 \\ -0.14} &	3.67\substack{+0.1 \\ -0.08} &	4.52\substack{+0.11 \\ -0.1} &	0.1384\substack{+0.0015 \\ -0.0015} &	n \\
1759.01 &	18.850009\substack{+2.2e-05 \\ -2.3e-05} &	1745.4661\substack{+0.0009 \\ -0.001} &	4.98\substack{+0.17 \\ -0.1} &	3.12\substack{+0.12 \\ -0.08} &	3.59\substack{+0.09 \\ -0.05} &	3.53\substack{+0.04 \\ -0.05} &	0.1147\substack{+0.0007 \\ -0.0005} &	n \\
1775.01 &	10.240554\substack{+1e-05 \\ -1e-05} &	1877.5645\substack{+0.0005 \\ -0.0005} &	8.9\substack{+0.1 \\ -0.09} &	8.05\substack{+0.14 \\ -0.13} &	3.66\substack{+0.04 \\ -0.03} &	3.66\substack{+0.04 \\ -0.04} &	0.0898\substack{+0.0005 \\ -0.0005} &	n \\
1776.01 &	2.799865\substack{+2.5e-05 \\ -3.2e-05} &	1871.498\substack{+0.0038 \\ -0.0037} &	1.19\substack{+0.09 \\ -0.07} &	1.22\substack{+0.09 \\ -0.08} &	1.93\substack{+0.19 \\ -0.18} &	2.5\substack{+0.06 \\ -0.05} &	0.0378\substack{+0.0004 \\ -0.0004} &	n \\
1778.01 &	6.527337\substack{+4.4e-05 \\ -4e-05} &	1876.0046\substack{+0.0025 \\ -0.0026} &	2.02\substack{+0.12 \\ -0.1} &	2.9\substack{+0.19 \\ -0.14} &	2.98\substack{+0.18 \\ -0.15} &	4.31\substack{+0.05 \\ -0.09} &	0.0726\substack{+0.0006 \\ -0.0006} &	n \\
1794.01 &	8.765566\substack{+7.3e-05 \\ -6.8e-05} &	1715.309\substack{+0.004 \\ -0.0033} &	2.3\substack{+0.11 \\ -0.09} &	3.3\substack{+0.17 \\ -0.14} &	3.92\substack{+0.13 \\ -0.13} &	5.11\substack{+0.12 \\ -0.1} &	0.0824\substack{+0.0009 \\ -0.0009} &	n \\
1797.01 &	1.039141\substack{+5e-06 \\ -3e-06} &	1900.238\substack{+0.0014 \\ -0.0016} &	1.31\substack{+0.06 \\ -0.05} &	1.47\substack{+0.07 \\ -0.07} &	1.82\substack{+0.06 \\ -0.06} &	1.89\substack{+0.03 \\ -0.04} &	0.0206\substack{+0.0002 \\ -0.0002} &	n \\
1797.02 &	3.645143\substack{+8e-06 \\ -8e-06} &	1902.8746\substack{+0.0014 \\ -0.0014} &	2.77\substack{+0.11 \\ -0.08} &	3.12\substack{+0.14 \\ -0.1} &	2.36\substack{+0.06 \\ -0.06} &	2.89\substack{+0.05 \\ -0.06} &	0.0476\substack{+0.0004 \\ -0.0005} &	n \\
1798.01 &	0.437815\substack{+1e-06 \\ -2e-06} &	1739.0717\substack{+0.001 \\ -0.001} &	1.62\substack{+0.08 \\ -0.06} &	1.4\substack{+0.07 \\ -0.06} &	1.28\substack{+0.04 \\ -0.05} &	1.19\substack{+0.02 \\ -0.02} &	0.0107\substack{+0.0001 \\ -0.0001} &	n \\
1798.02 &	8.021543\substack{+2.9e-05 \\ -3.3e-05} &	1741.5941\substack{+0.0022 \\ -0.0022} &	2.76\substack{+0.13 \\ -0.1} &	2.39\substack{+0.12 \\ -0.09} &	3.14\substack{+0.11 \\ -0.08} &	3.12\substack{+0.04 \\ -0.06} &	0.0744\substack{+0.0005 \\ -0.0006} &	n \\
1799.01 &	7.085754\substack{+8.4e-05 \\ -9.2e-05} &	1904.8305\substack{+0.0084 \\ -0.008} &	1.37\substack{+0.09 \\ -0.08} &	1.42\substack{+0.09 \\ -0.09} &	3.14\substack{+0.18 \\ -0.18} &	3.45\substack{+0.08 \\ -0.08} &	0.0707\substack{+0.0008 \\ -0.0008} &	n \\
1801.01 &	10.643985\substack{+2.7e-05 \\ -2.5e-05} &	1903.5423\substack{+0.0013 \\ -0.0014} &	3.51\substack{+0.11 \\ -0.11} &	1.96\substack{+0.08 \\ -0.08} &	2.7\substack{+0.06 \\ -0.05} &	2.67\substack{+0.04 \\ -0.04} &	0.0754\substack{+0.0005 \\ -0.0005} &	n \\
1807.01 &	0.549371\substack{+1e-06 \\ -1e-06} &	1900.4436\substack{+0.0004 \\ -0.0003} &	1.84\substack{+0.09 \\ -0.07} &	1.5\substack{+0.08 \\ -0.07} &	0.98\substack{+0.02 \\ -0.02} &	1.23\substack{+0.03 \\ -0.03} &	0.0122\substack{+0.0001 \\ -0.0001} &	n \\
1823.01 &	38.81359\substack{+3.3e-05 \\ -3.3e-05} &	1715.179\substack{+0.0004 \\ -0.0004} &	8.59\substack{+0.11 \\ -0.08} &	7.54\substack{+0.33 \\ -0.25} &	5.92\substack{+0.04 \\ -0.04} &	5.72\substack{+0.21 \\ -0.26} &	0.2118\substack{+0.0022 \\ -0.0023} &	n \\
1824.01 &	22.80853\substack{+6e-05 \\ -6.2e-05} &	1879.5468\substack{+0.0013 \\ -0.0012} &	3.1\substack{+0.09 \\ -0.07} &	2.74\substack{+0.09 \\ -0.07} &	4.11\substack{+0.06 \\ -0.05} &	4.51\substack{+0.07 \\ -0.08} &	0.1508\substack{+0.0011 \\ -0.0013} &	n \\
1836.01 &	1.772745\substack{+6e-06 \\ -5e-06} &	1929.5241\substack{+0.0018 \\ -0.0024} &	1.45\substack{+0.12 \\ -0.08} &	2.6\substack{+0.22 \\ -0.14} &	1.79\substack{+0.06 \\ -0.08} &	3.46\substack{+0.07 \\ -0.07} &	0.0309\substack{+0.0002 \\ -0.0003} &	n \\
1836.02 &	20.380831\substack{+2.5e-05 \\ -2.7e-05} &	1933.166\substack{+0.0008 \\ -0.0008} &	4.61\substack{+0.07 \\ -0.06} &	8.28\substack{+0.2 \\ -0.17} &	6.62\substack{+0.07 \\ -0.05} &	7.97\substack{+0.16 \\ -0.16} &	0.1574\substack{+0.001 \\ -0.0017} &	n \\
1842.01 &	9.573924\substack{+1.5e-05 \\ -1.5e-05} &	1933.3359\substack{+0.0009 \\ -0.0009} &	5.57\substack{+0.07 \\ -0.07} &	12.35\substack{+0.28 \\ -0.29} &	4.31\substack{+0.08 \\ -0.07} &	7.32\substack{+0.16 \\ -0.15} &	0.1002\substack{+0.0007 \\ -0.0008} &	n \\
1898.01 &	45.522288\substack{+6.1e-05 \\ -6.2e-05} &	1894.252\substack{+0.0007 \\ -0.0007} &	5.34\substack{+0.09 \\ -0.14} &	9.4\substack{+0.23 \\ -0.31} &	4.27\substack{+0.06 \\ -0.06} &	10.28\substack{+0.22 \\ -0.19} &	0.2687\substack{+0.0019 \\ -0.0027} &	n \\
2019.01 &	15.32222\substack{+0.000181 \\ -0.000172} &	1942.9799\substack{+0.0042 \\ -0.0054} &	2.8\substack{+0.13 \\ -0.1} &	5.33\substack{+0.26 \\ -0.21} &	7.57\substack{+0.24 \\ -0.26} &	7.67\substack{+0.26 \\ -0.13} &	0.1277\substack{+0.0038 \\ -0.0009} &	n \\
2045.01 &	9.077535\substack{+8.5e-05 \\ -8.4e-05} &	1765.5949\substack{+0.0014 \\ -0.0014} &	6.59\substack{+0.12 \\ -0.1} &	12.77\substack{+0.36 \\ -0.34} &	5.9\substack{+0.09 \\ -0.08} &	6.62\substack{+0.16 \\ -0.16} &	0.0927\substack{+0.0008 \\ -0.0019} &	n \\
2076.01 &	10.355307\substack{+1.4e-05 \\ -1.5e-05} &	1743.724\substack{+0.001 \\ -0.0009} &	3.19\substack{+0.12 \\ -0.07} &	2.77\substack{+0.14 \\ -0.1} &	3.24\substack{+0.05 \\ -0.04} &	3.44\substack{+0.09 \\ -0.12} &	0.0883\substack{+0.0007 \\ -0.0007} &	y \\
2076.02 &	21.015324\substack{+2.3e-05 \\ -2.4e-05} &	1748.6895\substack{+0.0006 \\ -0.0006} &	4.26\substack{+0.06 \\ -0.05} &	3.69\substack{+0.14 \\ -0.11} &	4.19\substack{+0.04 \\ -0.03} &	4.4\substack{+0.12 \\ -0.16} &	0.1415\substack{+0.0011 \\ -0.0012} &	y \\
2076.03 &	35.125611\substack{+9.6e-05 \\ -8.5e-05} &	1762.6658\substack{+0.0016 \\ -0.0019} &	3.95\substack{+0.2 \\ -0.27} &	3.43\substack{+0.21 \\ -0.25} &	2.97\substack{+0.12 \\ -0.09} &	5.2\substack{+0.14 \\ -0.18} &	0.1993\substack{+0.0016 \\ -0.0016} &	n \\
2088.01 &	124.730182\substack{+0.00061 \\ -0.000575} &	1769.6077\substack{+0.0031 \\ -0.0032} &	3.98\substack{+0.13 \\ -0.1} &	3.68\substack{+0.19 \\ -0.14} &	7.36\substack{+0.15 \\ -0.14} &	8.32\substack{+0.27 \\ -0.36} &	0.4722\substack{+0.0039 \\ -0.0044} &	n \\
2114.01 &	6.209837\substack{+0.000206 \\ -0.000197} &	2719.047\substack{+0.0011 \\ -0.0011} &	6.17\substack{+0.07 \\ -0.08} &	14.1\substack{+0.33 \\ -0.32} &	4.6\substack{+0.08 \\ -0.08} &	6.55\substack{+0.11 \\ -0.13} &	0.0753\substack{+0.0005 \\ -0.0006} &	n \\
2128.01 &	16.341418\substack{+0.000145 \\ -0.000128} &	1987.2651\substack{+0.0022 \\ -0.0022} &	1.71\substack{+0.06 \\ -0.05} &	2.09\substack{+0.09 \\ -0.08} &	4.7\substack{+0.14 \\ -0.11} &	5.22\substack{+0.14 \\ -0.13} &	0.1271\substack{+0.0016 \\ -0.0016} &	n \\
2145.01 &	10.261125\substack{+1.1e-05 \\ -1.2e-05} &	2013.2802\substack{+0.0006 \\ -0.0006} &	4.14\substack{+0.05 \\ -0.04} &	12.42\substack{+0.31 \\ -0.25} &	7.47\substack{+0.05 \\ -0.04} &	9.52\substack{+0.21 \\ -0.19} &	0.1107\substack{+0.0007 \\ -0.0009} &	n \\
\enddata
\tablecomments{Properties of 108 planets orbiting the 85 star TKS sample. Orbital period $P$, transit mid-point $t_{0}$ (given in BTJD = BJD - 2457000), transit duration $T_{14}$, and planet-to-star radius ratio $R_\textrm{p}/R_\star$ were measured from a final fit using \texttt{exoplanet} (\citealt{Foreman-Mackey2021}) and reweighted via importance sampling. The given values reflect median measurements with upper and lower uncertainties.  $R_\textrm{p}$ follows from $R_\textrm{p}/R_\star$ and $R_\star$. The expected duration of a centrally transiting object on a circular orbit $T_\textrm{circ}$ and semi-major axis $a$ are determined from Kepler's Third Law (\citealt{Winn10}). The last column shows a flag (yes or no) indicating whether or not observed TTVs are significant for a given planet. }
\end{deluxetable*}

\bibliographystyle{aasjournal}
\bibliography{adslib}

\clearpage

\end{document}